\documentclass[final]{siamltex}
\usepackage{graphicx,url,amssymb}

\def\ait{{A{\"\i}t-Sahalia }}

\title{Fitting Effective Diffusion Models to Data Associated with a ``Glassy Potential": \emph{Estimation, Classical Inference Procedures and Some Heuristics} \thanks{This
        work was partially supported by a Ford Foundation/NRC Fellowship}          }%

\author{Christopher P. Calderon \thanks{Department of Chemical Engineering, Princeton
        University, 
         Princeton, New Jersey
        08544 ({\tt ccaldero@princeton.edu}).}
        }

\begin{document}
\maketitle

\begin{abstract}
 A variety of  researchers \cite{yannisARXIV,klaus,hummer,yannisgear,dima,stu}
have successfully obtained the parameters of low-dimensional diffusion models  using the
data that comes out of atomistic simulations. This naturally raises a variety of questions about efficient estimation, goodness-of-fit tests, and confidence interval estimation.  The first part of this article
 uses maximum likelihood estimation (MLE) to obtain the parameters of
 a diffusion model from a scalar time series. I address
 numerical issues associated  with attempting to realize asymptotic statistics
 results with moderate sample sizes in the presence of exact and approximated transition
 densities. Approximate transition densities are used because the
 analytic
 solution of a transition density associated with a parametric diffusion model is often unknown.
 I am primarily interested in how well the deterministic transition
  density expansions of A{\"\i}t-Sahalia capture the curvature of the transition density in
 (idealized) situations that occur when one carries out simulations
  in the presence of a ``glassy" interaction potential.
   Accurate approximation of the curvature of the transition density
    is desirable because it can be used to quantify the
    goodness-of-fit of the model and to calculate asymptotic confidence intervals of the estimated parameters.
    The second part of this paper contributes
    a heuristic estimation technique for
    approximating a nonlinear diffusion model.
     A ``global"  nonlinear model is obtained by
     taking a batch  of time series and applying
     simple local models to portions of the data.
      I demonstrate the technique on a diffusion model with a
      known transition density
      and on data generated by the
      Stochastic Simulation Algorithm \cite{tau2}.

\end{abstract}



\pagestyle{myheadings}
\thispagestyle{plain}
\markboth{C. Calderon}{Estimating Effective Diffusion Models}

\section{Introduction}
Complicated systems are often approximated by overly simplified
models. A significant research effort has gone into attempting to
efficiently summarize the information contained in a complicated 
atomistic simulation with a low-dimensional effective model
\cite{yannisARXIV,klaus,hummer,yannisgear,dima,stu}.  Atomistic simulations contain many observables, an effective diffusion model aims at representing the salient features of the data in the drift component of a stochastic differential equation (SDE) and lumping the effects of the neglected details into the noise term. The appeal of
effective models stems from the fact that information contained in
the effective models can easily and quickly be extracted by
analytical methods or well-established numerical procedures. The
idea being that the ``truth" is contained in the atomistic
simulation, but the computational load required to get the
information is so large that the researcher has a difficult time
exploring all of the process parameters under study.

Before one attempts to wrap effective models around the output
of an atomistic simulation, a variety of assumptions need to be
made about the data and the parametric model. In this paper, I
obtain parameter estimates from the classical parametric
framework via maximum likelihood estimation (MLE).
Here the phrase ``classical parametric
framework" refers to the fact that one uses a single family of functions (of specified functional form)
for the drift and diffusion coefficient functions of the
diffusion SDE 
that depends only on a finite number of parameters. I make the
following assumptions about the atomistic system and the
parametric diffusion model: \\[.25cm]

\begin{itemize}
    \item A small set of order parameters \cite{yannisDIFFMAP,schulten} that accurately summarize
    the full atomistic system are identified and easily measurable.  The term ``order parameter" is used to refer to the observables modeled in the effective diffusion SDE. 
     \item The parametric model is uniquely identifiable.
     \item The \emph{exact} transition density of the parametric model has all of the regularity properties that make MLE attractive. Namely it is the \emph{asymptotically} most efficient estimator in the sense that the  properly normalized parameter distribution associated with the procedure converges to a normal distribution with the smallest asymptotic variance (in ``nonergodic" situations this assumption is relaxed). 
   \item The true parameter value admits a contiguous neighborhood and the process allows for a quadratic expansion of the log likelihood ratio \footnote{These terms are briefly introduced in section \ref{s_stattools}, consult van der Vaart \cite{vandervaart} chapters 5-8 for a clear detailed treatment}.
    \item The dynamics of the order parameters can be adequately described by a diffusion SDE.
    \item The drift and diffusion coefficient of the effective SDE are suitably smooth (to be more specific assume the functions are infinitely differentiable with respect to the parameters and state, but this can be relaxed substantially) and  the drift component of the SDE comes from the gradient of an effective potential \cite{yannisARXIV,dima}. Even if the order parameter being considered is governed by a glassy potential (this term is described in section \ref{motivation}), a smooth drift coefficient function  can be used to summarize key features of the free energy landscape \footnote{ For example the global minimum value of the smooth effective  free energy surface is the same as that of the more complicated glassy surface}.  
 
\end{itemize}

The first item is extremely important and is an active area of
research in our group \cite{yannisDIFFMAP}, but will not be
addressed in this article. The next two assumptions allow one to
trust the  parameter estimates  and allows one hope for checking
if some asymptotic results associated with MLE
\cite{jeganathan,vandervaart,white} hold for the sample
sizes used. The fifth item is briefly
addressed in the final application of the second part of this
paper where I estimate the parameters of a diffusion approximation of a
jump process (the model of the
reduction of nitric oxide on a platinum surface given in
\cite{makeev} is revisited). 
  The final issue is concerned with ``model misspecification" \cite{white},  dealing
  with this issue is important if the effective model is to be of any
  practical use. I first obtain parameter estimates assuming the final item holds,
  but in section \ref{s_good} classical techniques for testing this
  assumption \emph{a posteriori} are outlined.  Some nonparametric goodness-of-fit tests are  better
  suited for practical implementation \cite{diebold,hong}, but  
  classical tests are of interest because the tests used quantify how well a transition
  density approximation matches the
  curvature of the true density.  
  
  In order to use MLE one must have an approximation of the transition density of the diffusion process.  Unfortunately, for many SDE's the transition density is not available in closed-form so one must resort to some approximation of the density.  
The estimation techniques presented are intended for application to interesting atomistic systems (e.g. a time series that comes from a molecular dynamics simulation), but in this paper toy models are studied in order to systematically study the consequences of using the transition density expansions of \ait \cite{ait2,aitECO} in some large sample statistics applications.  Our research group has had success in applying some of these ideas to actual atomistic systems, but this paper's main concern is in determining exactly how much information can be extracted from the transition density expansions in controlled examples intended to mimic scenarios encountered in some multiscale applications.

The remainder of this paper is organized as follows: Section \ref{motivation} reviews the multiscale applications that motivated this study.  Section
\ref{s_model} lays out the model systems used. Section
\ref{s_stattools} outlines the techniques and estimation
tools used in the presence of ``contaminated" \footnote{This term is used to convey the fact that one knows that the true data does not follow the exact proposed parametric model}  data and transition
densities.  Section \ref{s_locmod}  outlines a simple estimation
procedure that can be used to study  multiscale systems that
satisfy the assumptions stated above. Section \ref{s_results}
contains the numerical results and discussion and the final
section gives the conclusion and outlook.

\section{Motivation}
\label{motivation}
A situation encountered often in spin glasses
\cite{pablo}, protein folding \cite{schulten}, and zeolites \cite{deem} is that of
a harmonic ``glassy" potential energy surface \footnote{This estimation strategy was developed in order to accurately measure the curvature of complicated free energy surfaces associated with atomistic simulations.  If the location(s) of the dominant free energy wells are known by theory or simulation methods, then the estimation methods shown  here can be used to measure the curvature at the well minima which can in turn be useful for getting information about transition pathways \cite{pittguy}. If one also has knowledge of where the saddles are and a protocol for starting meaningful simulations around the saddle point then the methods presented can also be used to determine the curvature of the unstable state points.   }. That is to say, when one looks at a free-energy surface from a distance, the shape of the free energy surface is roughly parabolic.  When one looks closely at the details of the surface however, one sees many bumps in the surface (see figure \ref{fglassyPOT}).
    In many applications, it is believed that the order parameter of the process ``funnels" its way
 down to the global minimum of the free energy surface \cite{schulten} in the long time limit.
  Current computational power does not
  always allow one to carry out an atomistic
  simulation long enough to observe such a phenomena because the order parameter can get trapped in a local minima.  Sometimes one
   can  reach the global free energy minima by increasing the temperature
   parameter of the simulation  \cite{frenkel,paralleltemperingDEEM}.
   When this occurs, the force binding
    the order parameter to the global minimum still has a ``bumpy" potential
    associated with it (but the magnitude of the bumps is smaller because of the new temperature scale). I refer to this hypothetical case as  situation I. 

     Another commonly encountered scenario is one where the
     order parameter is trapped in a free energy well which is not the global
      minimum of the surface.  Many applications require system information at low temperatures, ruling out the simple technique mentioned in the previous paragraph .     If the temperature of the system is so
      low that on the timescale of the atomistic simulation that the order
      parameter appears to be approximately  bound by a smooth (but not necessarily harmonic) potential in a
      neighborhood of the local minima, then I will refer to this case as situation II.

       Classic statistical mechanics models usually assume that the noise around the
       local minima is state independent. This assumption does not hold in a variety
       of interesting systems, \cite{dima,yannisARXIV} so in all models I consider there
       is state dependent noise in the process. 
The estimation techniques presented in this paper  deal with
both situations
described in the preceding paragraphs.

\section{Model Systems}
\label{s_model} To idealize situation I and II,  data is generated 
from two families of SDE's.  This first family  is meant to mimic
situation I and  has the form:
\begin{equation}
dX_t=\Big(\kappa \left(\alpha-X_t \right) +\beta \sin \big(
 \left( X_t-\alpha \right) \omega 2\pi \big) 
\Big) dt +\sigma\sqrt{X_t}dW_t
\end{equation}

The second family idealizes situation II and takes the form:
\begin{equation}
dX_t=\left(\kappa \left(\alpha-X_t \right) +4\gamma
\left(\alpha-X_t \right)^3 \right)dt+\sigma\sqrt{X_t}dW_t
\end{equation}

 The parameters are set to $\alpha=20$, $\kappa=\sigma=4$, and $\omega=\frac{1}{3}$ throughout.
The parameters $\beta$ and $\gamma$ take the values (0,15,60,200) and (0,$\frac{1}{400}$).
 I refer to these  cases as situations I A-D and situations II A-B respectively.
  The situation where $\gamma$ and $\beta$ are both zero is known as the Cox-Ingersoll-Ross (CIR)
  model and is one of the rare situations where an SDE with  nonlinear
   coefficients has an explicit solution \cite{ait2}.
   I use this example because it illustrates mean reversion, it demonstrates  state dependent noise
     and most
    importantly has an exact closed-form transition density which can be used to help determine why a
    transition density approximation is failing.

\begin{figure}
\includegraphics[angle=0,scale=.65]{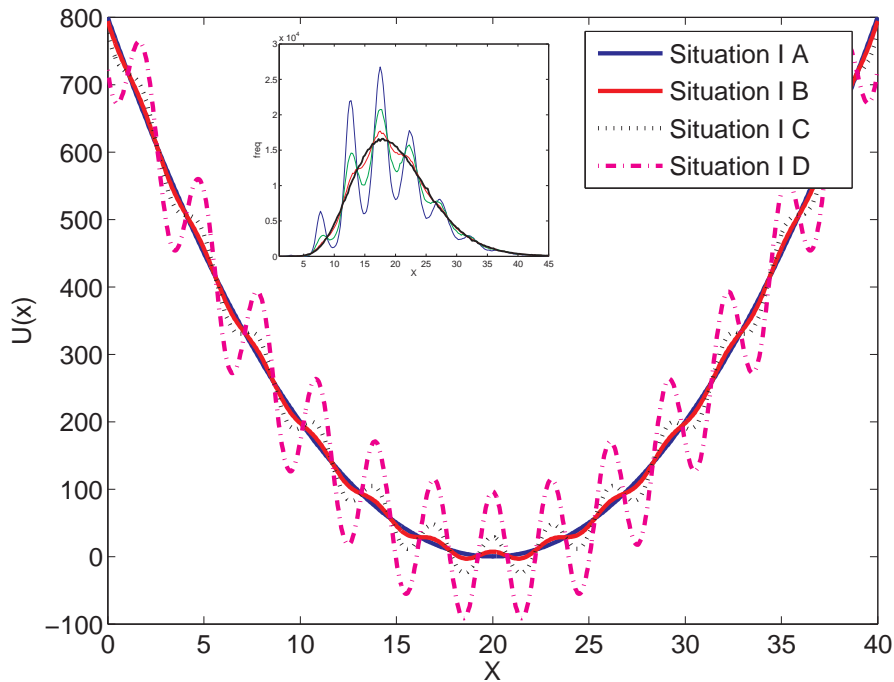}
\caption{Situation I Potential \small{ 
\emph{Potential energy function used to determine drift with $\beta=(0,15,60,200)$. The inset shows the empirically measured invariant distribution  for the
 four values of $\beta$ used (with obvious correspondence between the four cases); the distributions are shown only
   to give one an idea of
  how the different parameter values
  affect the long-term dynamics (MLE parameters only depend on the observation frequency). }} 
}

\label{fglassyPOT}
\end{figure}

\begin{figure}
\includegraphics[angle=0,scale=.65]{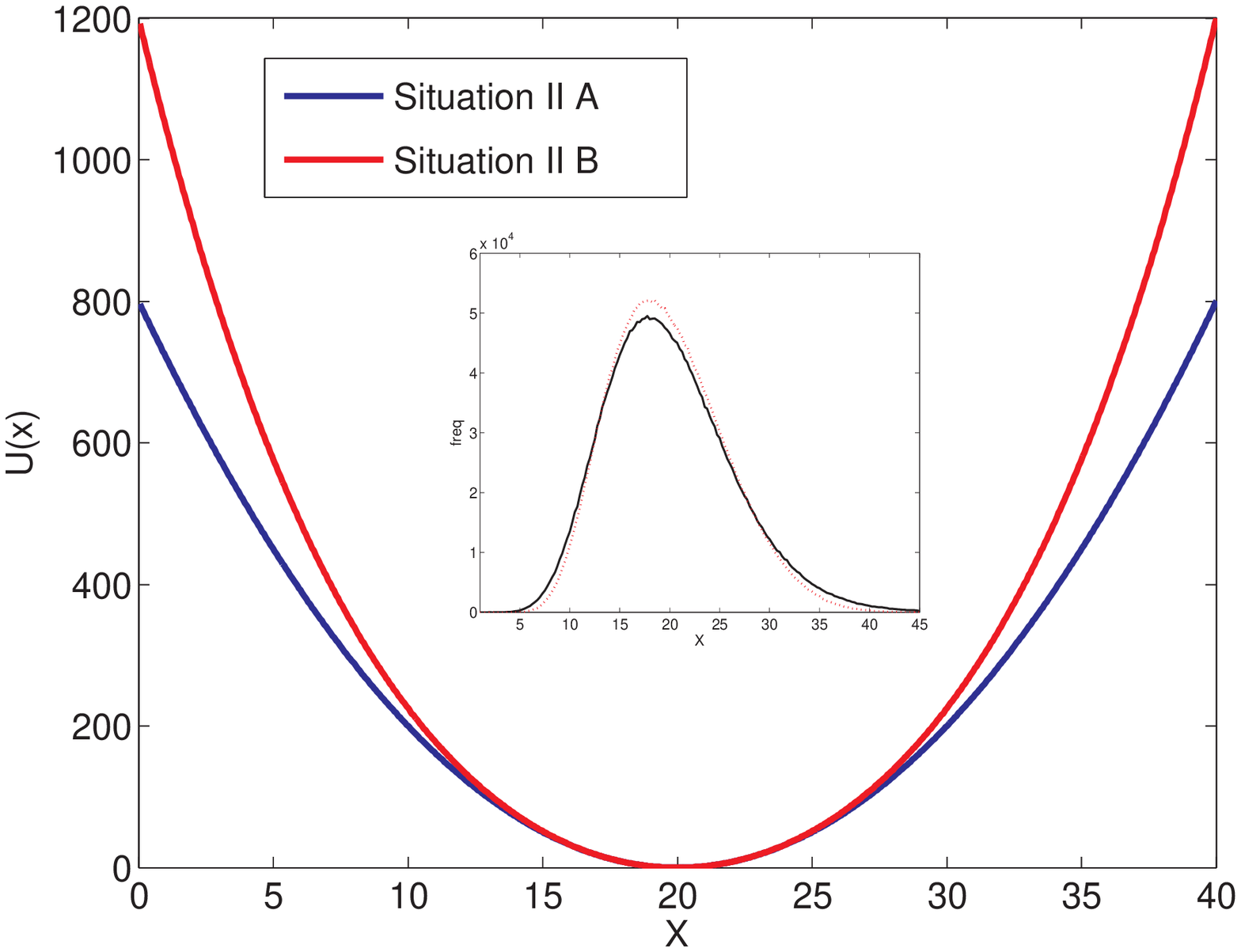}
\caption{Situation II Potential \small{
\emph{ Potential energy function used to determine drift with $(\gamma=0,\frac{1}{400})$ along with inset of corresponding invariants distributions (the narrower distribution corresponds to the nonlinear perturbation case)}}}

\label{fnonlinpot}
\end{figure}

Figures \ref{fglassyPOT} and \ref{fnonlinpot} plot the potential
energy surfaces for the cases studied.  In all cases, 
sample paths of the above processes are simulated using the explicit Euler scheme
\cite{kp} with a step size $\Delta t=2^{-9}$ given data starting from the invariant density associated with situation I A.   The data is observed
every $16^{th}$ step yielding a
constant observation window spaced by $\delta t_{obs}=2^{-5}$ time
units (in Situation I A-C  the data is also sampled every
$64^{th}$ step giving $\delta t_{obs}=2^{-3}$).  Each plot and
graph that are grouped together in this paper used the
same Brownian trajectories in order to simulate paths  (the only
difference between the sample paths is caused by the different
drift coefficients) in order to
reduce variation due to random number draws.

For the first part of this paper, the data generated by the
SDE's above is modeled by the CIR class:
\begin{equation}
\label{CIR} dX_t=\kappa(\alpha-X_t)dt+\sigma\sqrt{X_t}dW_t
\label{eq_cir}
\end{equation}

For the second part of this paper, the following parametric
family is used:
\begin{equation}
 dX_t=\Big(a+ b(X_t-X_o)\Big)dt+\Big(c+ d(X_t-X_o)\Big)dW_t
 \label{eq_myaffine}
\end{equation}
Where $X_o$ is user specified;  the
parameter vectors are estimated by  techniques associated with maximum likelihood in all cases.

The second terms in the drift coefficient of the data generating
processes are used to determine how robust the estimator is
against model misspecification \cite{white,millar,semipar}.
The perturbation terms are not modeled because it assumed that
their true (or approximate)
functional form are completely unknown and the interest is primarily in the  smooth
noise and mean reversion parameters ($\alpha,\kappa,\sigma$). Of course the presence of
these extra terms affect the estimation of the parameters, it is
shown in section \ref{results} that the effects induced by these
perturbation parameters affect things consistently  with what a
physicist of chemist would intuitively anticipate. The interesting
feature demonstrated in the  aforementioned section is
quantitatively how the MLE procedure carried out with various
transition density approximations respond to these  perturbation
parameters  in relation to the exact transition density.

 The reason for studying the first two
idealized model systems stems from a desire to 
\emph{carefully} numerically quantify how the MLE procedure with
approximated transition density performs in tasks beyond point parameter
estimation. The results obtained are of course specific to the
model and parameter values used, however one can always obtain the
parameters of a parametric diffusion model using observations from the particular system being
studied and then carry out an idealized set of tests
similar to the ones presented here by using established SDE path simulation techniques \cite{kp}.

The final model presented is one for the
reduction of nitric oxide ($NO$) by hydrogen gas on a platinum  (Pt) surface. The
mechanism is as follows \cite{yanniscatalyst}:
\begin{eqnarray}
NO  \stackrel{k_1}{\rightarrow} NO^\dagger \nonumber \\
NO^\dagger  \stackrel{k_2}{\rightarrow}  NO \nonumber \\
H_2 + NO^\dagger \stackrel{k_3}{ \rightarrow} \frac{1}{2}N_2 + H_2O \nonumber \\
\end{eqnarray}

$NO{^\dagger}$ represents $NO$ absorbed onto the Pt surface;  this mechanism is used with Gillespie's Stochastic Simulation Algorithm (SSA) \cite{tau2} technique in order to construct stochastic evolution rules for the amount of $NO$ in the system at any given time. This
model is used because it exhibits nonlinear mean reversion with state dependent noise. The jump
process is known to converge weakly to a diffusion with a cubic drift
term as the system size parameter (denoted by $N_{molecules}$) increases \cite{tau2,kurtzbook}. This yields  another ``situation II"  type scenario, but now
the data generating process is not a genuine diffusion  \footnote{It should be pointed  out that some approximations of the above process by a diffusion model \cite{tau2} use as many Brownian driving terms as there are elementary reaction steps; our black-box approach only uses one Brownian term for each state component (in this paper the noise contribution  of the individual reaction events are lumped into a single noise term) .}. The parameter $k_3$
is set to a numerical value of 4, the other
model parameters used are given in \cite{makeev}.  In the final
part of this paper, the parameters of the assumed model are extracted   in the ``small
molecule" case ($N_{molecules}=3600$). This value is chosen
because simple visual inspection indicates that the process has not yet converged to a diffusion (
$\delta t_{obs}=2^{-5}$) with
this molecular population size, so the exact functional form
of the diffusion model is unknown. A ``global" \footnote{Actually one can only estimate the function for state points visited, making global a slight misnomer (hence the quotes). One is always free to extrapolate the coefficient functions and get a genuine global diffusion approximation if one somehow knows beforehand that the state points in the time series are  adequately representative of the entire portion of phase space having significant probability mass in the infinite time limit.       
\label{globaldisclaimer}} estimate of the diffusion approximation of the actual process is obtained
using techniques presented in the second part of this paper and
 the invariant density of the obtained nonlinear diffusion
model is compared to that of the actual SSA process in section \ref{s_results}.

\section{Statistical Tools}
\label{s_stattools}

 In the beginning of this
 section, 
 some classical statistical tools relevant to this study are outlined.
 The tools are only briefly defined,
 references are given throughout which comprehensively describe the details of the theory applied. The tools below are applied to the estimation of the parameters associated with both stationary and ``nonergodic" 
 \footnote{ This term is intended to describe 
situations where the parameter distribution associated with an
estimation scheme is not asymptotically normally distributed (with
a deterministic covariance matrix); this can occur if
the sample size is itself random or if the time series is
nonstationary \cite{basawa}.} time series.

MLE's importance \footnote{In practice one usually deals with a quasi-maximum likelihood estimate.  The difference between the two is described in section \ref{s_good}} stems from the fact that one needs some kind of generic metric by which to judge a wide class of parametric models by.  In situations where the underlying transition density satisfies a set of regularity assumptions \cite{vandervaart}, it is very appealing because it provides a consistent estimator with the minimum asymptotic variance. If one has a reliable estimate of the underlying transition density, then one can sometimes (in the stationary ergodic distribution case \cite{jeganathan}) determine the asymptotic parameter distribution with a simple deterministic integral \cite{hamilton,lecam00}.  One can also generate test statistics based on output of the MLE procedure which can be used to asses the goodness-of-fit of the parametric model \cite{white}. 


Next, an optimal simple hypothesis test is introduced.  Specifically, the
transition density expansions are used in order to create the
Neyman-Pearson test statistic \cite{bickel}.  A simple hypothesis
test is useful when one wants to test the statistical  significance  of the magnitude of the changes in
effective model parameters when one adds more features to the
underlying  atomistic simulation (e.g. one would like to determine
if the changes in the effective model are significant when the
parameters are estimated from the output of atomistic simulations that use a potential with and without electrostatic
interactions).

It is already known that the simple models that are wrapped around the
data do not faithfully represent the \emph{exact} system dynamics.
In section \ref{s_good}, methods that can be used to
quantify how closely the proposed parametric models represent the data are reviewed.  In section
\ref{s_laq}, a heuristic method that  can be used in the
nonergodic case for obtaining parameter uncertainty estimates is presented.

\subsection{Maximum Likelihood Basics}
\label{s_mlebasics}
In order to avoid
technical complications,
 it is assumed throughout that the \emph{exact} distribution  associated with the parametric model admits a
density whose logarithm is well defined almost everywhere and the
logarithm of the density is continuously twice differentiable.  The principal
of maximum likelihood is based on maximizing the following
integral with respect to the parameter $\theta$:
\begin{equation}
\int\limits_{\Omega}{f(\mathbf{x;\theta})\log\Big(f(\mathbf{x};\theta)\Big)d\mathbb{Q}}
\label{eq_infoint}
\end{equation}
In the above equation, $\mathbb{Q}$ corresponds to the measure of the underlying probability space, $f(\mathbf{x};\theta)$ corresponds to
the Radon-Nikodym derivative  (consult \cite{kl,white}) of the law of the random variable $\mathbf{x}$ with respect to the
underlying probability space and $\mathbf{x}$ corresponds to a
discretely sampled time series (of finite length = $M$).  Assume that  if the measure of a set under $\mathbb{Q}$ is zero it implies that the measure of the set under $\mathbb{P}_{\theta}$ is also zero (the phrase ``$\mathbb{P}_{\theta}$ is absolutely continuous with respect to $\mathbb{Q}$" is used to describe this situation).  Let $\mathbb{P}_{\theta}:=\int \limits_{\Omega} f(\mathbf{x};\theta)d\mathbb{Q} $.   For discrete
Markovian models, $f(\mathbf{x};\theta)$ can readily
calculated by the following formula \cite{hamilton}:

\begin{equation}
f(\mathbf{x};\theta)=f(\mathbf{x_0})\prod \limits_{n=1}^{M-1}
f(\mathbf{x_{n}}|\mathbf{x_{n-1}};\theta)
\end{equation}

In the above equation
$f(\mathbf{x_{n}}|\mathbf{x_{n-1}};\theta)$
 represents the conditional probability (transition
density) of observing $\mathbf{x_{n}}$ given the observation
$\mathbf{x_{n-1}}$.  In practice,  one usually takes a finite sample of data and
presents this data to a Monte Carlo scheme that is meant to approximate the integral in
equation \ref{eq_infoint} and finds the parameter values that yield
the maximum value.  In  what follows,  the function
below is  referred to as the log likelihood function (assume throughout that $\mathbf{x_o}$ has a Dirac distribution) :
\begin{equation}
\mathcal{L}_{\theta}:=\sum\limits_{i=1}^{M}
\log\Big(f(\mathbf{x_{i}}|\mathbf{x_{i-1}};\theta)\Big)
\label{eq_loglikelihoodfuncdef}
\end{equation}

  Under our assumptions one has \footnote{This actually holds under less stringent regularity assumptions  \cite{vandervaart}} the following:
\[
\sqrt{M}(\theta-\hat{\theta}) \stackrel{\mathbb{P}_{\hat{\theta}}}{\Longrightarrow} N(0,\mathcal{F}^{-1})
\]

Where in the above $\hat{\theta}$ is the ``true" parameter of the
model; $\theta$ represents the parameter estimated with a finite
time series of length $M$; $\stackrel{\mathbb{P}_{\hat{\theta}}}{\Longrightarrow}$ denotes convergence in
distribution \cite{vandervaart,hamilton} under $\mathbb{P}_{\hat{\theta}}$ ;  $N(0,\mathcal{F}^{-1})$
denotes a normal distribution with mean zero and covariance matrix
$\mathcal{F}^{-1}$.  For a correctly
 specified model, $\mathcal{F}$ \footnote{I will adhere to common convention and call this the Fisher information matrix} can be estimated in
 a variety of ways  \cite{white,lecam00}. 
 Various conditions can be tested to see if asymptotic
 results are relevant for the finite sample sizes used \cite{vandervaart,lecam00}. In this article sample
 sizes are moderate, but good agreement with some classical asymptotic
 predictions \cite{white} is observed.  When a closed-form transition density is in hand one can deterministically calculate $\mathcal{F}$ in the stationary ergodic case 
  \footnote{In the stationary ergodic case, if one has a time series ($\mathbf{x_1},...,\mathbf{x_M}$) then one can ignore the initial distribution in the ``infinite M" limit.  If the state is $n$-dimensional, then one can approximate the Fisher information by a $2\times n$ dimensional deterministic integral $\left(\mathcal{F} \approx \int  \frac{\partial
log(f(x|x_{o};\hat{\theta}))}{\partial \theta}
\frac{\partial log(f(x|x_{o};\hat{\theta}))}{\partial
\theta}^T f(x_|x_{o};\hat{\theta})dxd\pi(x_o) \right)$, where $d\pi(x_o)$ is the invariant distribution of the process (which in the scalar case can usually be
readily calculated in closed-form from the coefficients of the parametric SDE \cite{hong}).  The difficulties
encountered in the nonstationary time series situation is analogous to the situation of using the Metropolis
algorithm to sample phase space \cite{frenkel}; in principle a
deterministic integral could be evaluated, but with current
computational power quadrature is not possible due to the high
dimensionality of the problem}.

In practice, maximum likelihood does fail spectacularly for some
simple models because of singularities that can be observed with
the log likelihood function \footnote{Recall this term implies a
finite sample size approximation to equation \ref{eq_infoint}}.
 The classical example is the following:  one
assumes that a distribution is a mixture two Gaussians whose
 variance $\in (0,\infty)$. Then the finite sample MLE fails to exist in this simple case (see  \cite{bickel,vandervaart} for a more in depth discussion)
\footnote{ In practice, one can partially remedy this situation by finding a local minima using a variant of the
technique outlined in section \ref{s_laq}, but the new comer to MLE
should be aware that some failures of MLE \cite{lecamrev} are
not as easy to remedy, especially when one has data that
does not come from the assumed parametric model class}.

The aforementioned point brings us to an important observation
contained in this paper.  It is well known that
 the transition density
associated with a diffusion SDE can be solved through a
corresponding PDE (via the Kolmogorov equations \cite{aitECO}). One
can often prove that the density of the process has the regularity properties needed to make \emph{exact} MLE successful (\emph{a priori} regularity bounds associated with the log likelihood function is trickier) via techniques of harmonic
analysis \cite{nualart}, but many useful properties determined by analytic techniques are only applicable to the
\emph{exact} transition density.  Approximations to the transition density are needed in cases where the transition density is not available in closed-form. The approximations 
provided by A{\"\i}t-Sahalia's Hermite expansion \cite{ait2,aitECO} are 
useful for obtaining point parameter estimates, can capture the
curvature of the transition density enough to approximate
parameter distributions, and are sometimes accurate enough to
create test statistics needed for some hypothesis tests. However,
when one uses derivatives of the expansion for creating Wald or
Rao \footnote{These test are concerned with using the
Fisher information matrix as a normalizing matrix to create a
$\chi^2$ statistic.  Both tests can be used to construct
confidence ellipsoids around  parameter estimates
\cite{vandervaart,hamilton,bickel} } test statistics \cite{bickel}
it can introduce spurious singularities which complicates
squeezing all of the information that is theoretically possible
from MLE.
I use the ``Euler" estimator as a crude transition density approximation  to demonstrate some points in this paper   \footnote{This estimator is motivated by the Euler SDE simulation path technique \cite{kp}. One assumes that for a given observation pair $(\mathbf{x_n},\mathbf{x_{n+1}})$ that a normal distribution can be used whose standard deviation is given by the diffusion coefficient evaluated at $\mathbf{x_{n}}$ times the square root of the time between successive observations ($\sqrt{\delta t}$) and the conditional mean is given by applying the  deterministic explicit Euler scheme to the drift coefficient given $\mathbf{x_{n}}$}.  It is well known that the Euler estimator is significantly biased \cite{lo}, this fact helped to initiate a flood of transition density approximations in recent years \cite{ait2,aitECO,aitextension,bibby,gallant,smle} (just to mention a few).

\subsection{Optimal Binary Alternative Hypothesis Testing: The Neyman-Pearson Lemma}
\label{s_np} In this section it is assumed that one has a data set and
two parameter vectors. Assume that one parameter vector is the
``null hypothesis" and the other parameter vector is the
``alternative". The type I error probability is defined as the
probability of rejecting the null hypothesis when in fact the null
is true (classically denoted by $\alpha$).  The power of a test
against the alternative is the probability of rejecting the null
hypothesis when in fact the alternative is true. An optimal test statistic can be
found  which for a specified $\alpha$ maximizes the power
\cite{bickel}.  To create the test statistic define the likelihood ratio by
$L:=\frac{f(\mathbf{x};\theta_{Alt})}{f(\mathbf{x};\theta_{Null})}$;

one rejects the null if $L>\beta^{NP}$ where $\beta^{NP}$ is a scalar value that allows the equality
$\int\limits_{L>\beta^{NP}}f(\mathbf{x},\theta_{Null})dx=\alpha$

In the context of multiscale systems computations, this test is really only practical for stationary ergodic distributions (or for order parameters that are trapped for the duration of a simulation in a local free energy minima), but it provides us with insight as to how well the transition density captures the likelihood ratio of nearby parameter points.  Applications  shown later require a highly accurate approximation of the likelihood ratio. 


\subsection{Goodness-of-fit and Model Misspecification}
\label{s_good}
 It is possible to test if the log likelihood function is consistent with the proposed model structure by testing the following condition \cite{white}:
\[
\mathcal{F}_{Hessian} :=\frac{\partial^2 \mathcal{L}}{\partial \theta^2}=-\mathcal{F_{OP}}
\]

Where the Hessian is evaluated at $\hat{\theta}$ and $\mathcal{F_{OP}}$ is defined by:
\[
\mathcal{F_{OP}}:=\frac{1}{M}\sum\limits_{i=1}^{M} \frac{\partial
log(f(x_{i}|x_{i-1};\hat{\theta}))}{\partial \theta}
\frac{\partial log(f(x_{i}|x_{i-1};\hat{\theta}))}{\partial
\theta}^T
\]
In the above, the superscript $T$ denotes the transposition operation, $\mathcal{F_{OP}}$ is the
``outer product" matrix.  For correctly specified models, both $\mathcal{F_{OP}}$ and  $-\mathcal{F}_{Hessian}$ \cite{hamilton}
are valid estimates of $\mathcal{F}$. When a
 closed-form expansion is in hand,
 these quantities are easily
 computable after the optimal parameter is located.

For real data it is overly optimistic to expect to be able to
exactly parameterize the density of the process with a Euclidean
parameter vector. However, in some situations it is meaningful to
attempt to project the data onto the proposed model structure
\cite{semipar,white}  yielding a Quasi-Maximum Likelihood
Estimator (QMLE). When the true density does not lie in the
proposed parametric model class, one can still maximize the
integral in equation \ref{eq_infoint} yielding the estimator that
minimizes the Kullbeck-Leibler distance \cite{kl,millar}. Under
assumptions laid out in \cite{white}, this QMLE  converges to a normal distribution in the infinite sample
limit:
\[
\sqrt{M}(\theta-\hat{\theta}) \stackrel{\mathbb{P}_{\hat{\theta}}}{\Longrightarrow}  N(0,\mathcal{C})
\]
The matrix $\mathcal{C}$ ($:=\mathcal{F}_{Hessian}^{-1} \mathcal{F}_{OP} \mathcal{F}_{Hessian}^{-1}$)
replaces $\mathcal{F}^{-1}$.
This fact can be used to test the goodness-of-fit given the data and the optimal parameter vector
via the Rao or Wald test statistic \cite{white,bickel}.
The methodology laid out in \cite{white} is comprehensive and powerful, but in order to develop techniques which can be accessed by the diverse audience involved in multiscale modeling, algorithms available in standard packages/libraries (MATLAB,IMSL) are employed .

\subsection{Le Cam's Method and Likelihood Ratio Expansions (LAQ)}
\label{s_laq}
Lucien Le Cam was a major contributor to a variety of important asymptotic statistics results \cite{lecam60,lecam86,lecam00}; one of his major contributions was concerned with Locally Asymptotic Quadratic (LAQ) expansions of the log likelihood ratio (llr).  Denote the llr symbolically  by $\Lambda_{h,M}(\theta)$.  It is given (for discrete Markovian models) by:
\[
\Lambda_{h,M}(\theta)=\mathcal{L}_{\theta+\delta_M h}-\mathcal{L}_{\theta}
\]
Here $M$ is again the length of the times series
; $h$ is a vector perturbation with the same dimensionality as $\theta$;
$\delta_M$ is a matrix that scales the perturbation
and $\mathcal{L}_{\theta}$
denotes the llr evaluated at $\theta$; for later use let $h_M:=\delta_M h$. When
the assumptions behind LAQ hold \cite{jeganathan,lecam00}, one
can asymptotically approximate the above statistical experiment by a
normal limit experiment \footnote{See \cite{vandervaart,lecam00} for a
clear introduction to this topic, consult \cite{jeganathan} for
an application to time series analysis}. If one is in a
neighborhood of the true parameter ($\hat\theta$)  the LAQ
conditions imply
\[
\Lambda_{h,M}(\hat{\theta}) -(h_M^TS_M(\hat{\theta})-\frac{1}{2}h_M^TW_M(\hat{\theta})h_M)
\]
  
converges (in $P_{\hat\theta}$ probability) to 0.
In the above expressions, when $\Lambda_{h,M}({\theta})$ is
twice differentiable with respect to $\theta$,
$S_M(\theta)$ and $-W_M(\theta)$ play the role of
the first and second derivative (respectively) of the llr \cite{jeganathan} \footnote{ This
view is convenient for giving an intuitive interpretation, but a classical Taylor expansion of
$\Lambda_{h,M}(\tilde{\theta})$  may fail to exist; however the
LAQ ``derivative"  can still be defined} with respect to the parameter vector.

The contiguity  condition is a generalization of the concept of absolute continuity \cite{halmos}; it allows one to determine the asymptotic distribution of ``nearby" experiments which is sometimes useful for constructing hypothesis tests. Denote the true parameter by $\hat\theta$ and let  $\tilde\theta$ be contained in a 
$\delta_M  -$ neighborhood of $\hat\theta$.  If contiguity exists in the experiment, it \cite{vandervaart} implies the following limit distribution: 


\begin{equation}
\label{eq_laqN}
\Lambda_{h,M}(\tilde{\theta}) \stackrel{\mathbb{P}_{\hat{\theta}}}{\Longrightarrow} N\Big(-\frac{1}{2}h_M^TW_M(\tilde{\theta})h_M,h_M^TW_M(\tilde{\theta})h_M\Big)
\end{equation}

The llr is of interest to us because it provides a method for taking a classical parametric model and producing a scalar random variable which has the above limit distribution in the LAQ case (the practical utility of this theory is realized if the above normal distribution can be used to reliably approximate the more complicated distribution of the llr associated with the proposed parametric model for moderate sample sizes \cite{lecam86} ).

 There are many other important implications of LAQ,  the  method is only used in this paper to quantify the uncertainty associated with a nonergodic time series.  In the stationary ergodic case the matrix  $W_M(\hat\theta)$ coincides with deterministic  quantity $\mathcal{F}$ shown earlier, for nonergodic models the matrix is itself a random variable.   I repeat the construction in Le Cam chapter 6 \cite{lecam00} here in order to show how one uses the llr in order to estimate $W_M(\hat{\theta})$ (which can be used to roughly approximate the variance of the limit parameter distribution).  To simplify the situation,  assume that one is already within a $\delta_M$ neighborhood of $\hat{\theta}$ (this neighborhood is centered at $\tilde{\theta}$  ) and set  the matrix  $\delta_M=\frac{1}{\sqrt{M}}Id$ where $Id$ is the identity matrix.  Suppose $\theta \in \mathbb{R}^k$, then one takes a basis set of $\mathbb{R}^k$  (denote this set by $\{ b_1,\dots,b_k  \}$) and evaluates:
\[
\Lambda_{h,M}(\tilde{\theta}+\delta_M(b_i+b_j))
\]
for $i,j=1,\dots,k$;   the components of $W_M(\tilde{\theta})$ are estimated by

\begin{equation}
\small
W_M(\tilde{\theta})_{ij} \approx \frac{-[\Lambda_{h,M}(\tilde{\theta}+\delta_M(b_i+b_j))-\Lambda_{h,M}(\tilde{\theta}+\delta_M(b_i))-\Lambda_{h,M}(\tilde{\theta}+\delta_M(b_j))]}{ b_i b_j}
\label{eq_laqW}
\end{equation}

Under the LAQ assumptions, one can claim that  $W_M(\tilde\theta) \stackrel{\mathbb{P}_{\hat{\theta}}}{\Longrightarrow} W_M(\hat\theta)$ as $M$ tends to infinity .  The theory shown here (developed by Le Cam) is purely an
asymptotic one, however if one observes that the relation in
equation \ref{eq_laqN} holds, it gives one more confidence when equation
\ref{eq_laqW} is employed to approximate the quadratic expansion of the llr.  In the situation where $W_M(\tilde{\theta})$ is
random (as it will be in some of the applications presented), one
needs to repeat this procedure for a variety of paths \footnote{ In practice one may not have enough paths from an atomistic 
simulation to realize asymptotic statistics results.  If a diffusion model is truly valid,  one is always free to use the parameters obtained from a small sample in order to  simulate additional paths \cite{kp}, then use this information to create asymptotic error bounds associated with ``perfect" data.}  
and then
take expectations in order to get a rough estimate of the
asymptotic parameter distribution associated with the collection of sample paths.  The estimation methods that use the LAQ expansions are purely heuristic, the theory developed above is proved in great detail for independent random variables; asymptotic statements about the llr associated with general Markov chains are harder to make \cite{lecam86,rieder}.  The potential applications of these types of ideas in multiscale applications are merely shown via numerical results in this paper.    

\section{Local Polynomial Diffusion Models}
Thus far, the main  concern has been approximating the transition
density and quantifying the goodness-of-fit of our simple models
to the data. For the remainder of this section assume that the
data can adequately be described by some arbitrary nonlinear
effective SDE.  Recall, I am working under the assumption that even if the true
underlying potential energy surface is rugged, a smooth model of the drift and diffusion coefficients can
be used to reliably approximate certain features of the free
energy surface.
 In statistical mechanics, sometimes a limit theory exists for
 what the large sample system's
 effective dynamics converge to \cite{lebowitz}.
 Unfortunately, for many interesting systems,
 the functional form of the drift and
 diffusion coefficients  are completely unknown.
 Our research group has used the label
 ``equation free methods" \cite{yannisgear}
  to describe a set of numerical procedures that have been
  designed to deal with a situation where one has a
  simulation protocol that is believed to contain useful
  information and the information in the simulation is
  believed to be describable by an effective equation (with smooth coefficients),
  but the equation is unavailable in closed form.
  The early versions of this procedure used least
  squares approximation in order to get derivative
  information.  Within the last couple of years,
  our group has extended the estimation procedure to
  match local linear SDE models (both vector
  and scalar) to
  the output of simulation data.
  The parametric models 
  proposed are of the
  type given in equation \ref{eq_myaffine}.
  MLE and QMLE are greatly facilitated by the
  deterministic likelihood
  expansions developed
  by A{\"\i}t-Sahalia \cite{ait1,ait2,aitextension} \footnote{ Deterministic expansions are  useful because one can quickly carry out   parameter optimization and can easily evaluate the functionals needed to carry out
  variations of MLE in situations where MLE fails \cite{lecam00,jeganathan}.  Another obvious advantage is that in the case of stationary ergodic distributions one can easily carry out a deterministic quadrature and determine the asymptotic parameter distributions in situations where the parameter distributions converge to a normal random variable}.

    Conceptually, I am just taking advantage of the fact that the accurate \ait Hermite expansion allows us to generalize the piecewise-polynomial (pp) concept to 
    diffusion SDE models.  Since I posit the existence of a smooth underlying effective diffusion model,   the drift and diffusion coefficient functions are fit locally to linear models.  One is free to use a fairly broad class of polynomials in the scalar case with a variant of A{\"\i}t-Sahalia's expansion \cite{aitextension}, but our future work is concerned with  applying the techniques in this paper to the vector case.  In that situation, a higher order polynomial of unknown vector functions is not practical from an estimation standpoint.  Our experience has shown that if one wants to accurately capture the mean reversion portion of a nonlinear effective model that it usually becomes necessary to model the state dependence of the noise (  this point is illustrated in section \ref{s_reslocal}). 
    In statistical terms, I am simply just finding the QMLE estimate of
    the parametric model shown in equation \ref{eq_myaffine}.
However a variety of practical questions arise: \\[.25cm]

\begin{itemize}
    \item How does one determine if the estimated model is meaningful?
    \item In the case where an approximate transition density is used to estimate parameters, how sensitive is the QMLE projection to the quality of the approximation?
    \item How does one choose the size of the neighborhood where a local linear model is valid?
    \item How does one piece together the local models in order to create a global nonlinear effective diffusion model?
\end{itemize}

The first   issue is readily handled by the theory discussed
earlier. The second issue is concerned with semiparametric
estimation and robust statistics. These are important and
 active area of
 research in
 statistics \cite{semipar,vandervaart} (and in the future could make great contributions to multiscale modeling),
 but results are currently difficult (for this author)
 to translate into practical and reliable numerical methods associated with the ideas laid out here.
 In section \ref{s_mlebasics} it is shown
 that the true model is not too sensitive to
 ``contaminated data" and this quality is
 retained by the expansions of A{\"\i}t-Sahalia,
 however is not by the overly crude Euler estimator. 

\begin{figure}
\includegraphics[angle=0,scale=.65]{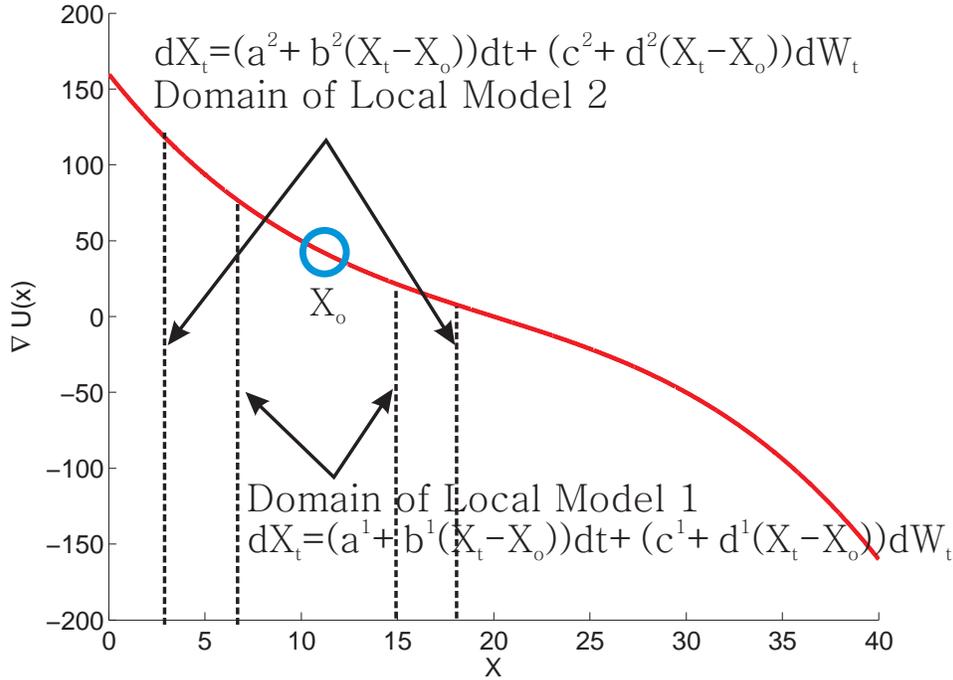}
\caption{LAQ Screening Method Illustration Applied to Situation II B Data \small{
\emph{ The circle  corresponds to a state point where a parametric model was obtained using that particular``$X_o$" in the parametric model given in equation \ref{eq_myaffine}.  For smooth models the quality of the linear approximation depends on  the neighborhood size as well as the curvature of the drift and diffusion coefficients (latter not shown).  Once the data is collected, one is free to vary $X_o$ and/or the neighborhood size where the local SDE is assumed to model the data.  } }
}

\label{hitLEVillustration}
\end{figure}

A simple estimation procedure is well-equipped to handle the third issue.  The  method  requires the user to obtain a batch of trajectories from a simulation.  In order to carry out a classical estimation procedure, one must create a partition of state space where a local linear diffusion model is an accurate representation of the underlying nonlinear diffusion (see figure \ref{hitLEVillustration}).  Once the data set is collected, one is free to make as many partitions as desired.  The simple idea is to only present observation pairs ($\mathbf{x_n},\mathbf{x_{n+1}}$) to the log likelihood function that have ``$\mathbf{x_n}$"  within the selected neighborhood.  The neighborhood size must be chosen to be small enough that the QMLE parameters estimates of the pp SDE model are statistically meaningful \footnote{ The goodness-of-fit techniques presented in the previous section can check this, however in practice this procedure greatly benefits from modern nonparametric techniques \cite{hong,diebold}} and large enough to contain enough samples to obtain the desired parameter accuracy (and have asymptotic limits remain useful for the sample size).  The main difficulty that the method faces is determining a neighborhood yielding  a satisfactory compromise between the two aforementioned factors.   This procedure is greatly facilitated by the equations of a limit process if one is known.  Otherwise one must use numerical experimentation to attempt to determine how smooth the underlying coefficient functions are (recall the assumption that a simple low-dimensional model exists).  In multiscale modeling we have the convenience of controlling the initial conditions of the (assumed meaningful) atomistic simulation making this type of pp SDE modeling more practical (otherwise one can only estimate parameters around state points visited frequently by the sample path).   Note that this method makes the time series length itself a random variable.  In addition, many applications of this idea lead to estimation of a nonstationary time series.  It has already been noted earlier that it is computationally difficult to calculate the Fisher information matrix in these circumstances. These facts indicate that the LAQ likelihood ratio approximation might be useful.  It should be noted that optimal tests rarely exist in this type of situation, however in the estimation literature some heuristic techniques have been recommended \cite{basawa,feigin}.  The technical details of these methods require some fairly specialized arguments and due to this author's ignorance of  both recent developments and the more important aspects of the theory, goodness-of-fit tests are avoided in this situation
 \footnote{Before proceeding, I would like to explicitly point out that if one has an initial distribution of points clustered near each other initially and as time proceeds the ensemble mean slowly ``funnels" its way down to global or local minima, then the problem is slightly easier because one only needs to determine the boundary  of the ``right hand end points" of the collection of time series.  The point  stressed is that data only needs to be collected \emph{once} and a statistically analyzed on or off-line (the deterministic expansions of \ait make the former a possibility)}. 

 In regards to the final item, I merely present two  simplistic methods of piecing  together our local models in section \ref{s_locmod}.  I also demonstrate  that the LAQ expansion can give a useful quantification of parameter uncertainty for our toy models.  The information contained in an LAQ approximation can be used in more sophisticated ``matching" schemes.  Again this area is well outside of this author's research area; if the problem at hand is of any real significance one should consult \cite{deboor,fan,wahba} for modern matching  techniques, but at some stage one will almost certainly have to appeal to some form of heuristics (a.k.a.,  ``the art of numerical computation").  
\label{s_locmod}
\subsection{Relevance to Multiscale Numerical Methods}
The details of  the tools used may obscure their utility in multiscale modeling, so   the connection is  summarized here.  The computational load to carry out a full simulation is too great, so a diffusion approximation is made which hopes to capture the relevant information in the underlying model.  Goodness-of-fit tests are needed to quantify the quality of the approximation.  If the model is found to be statistically acceptable, then one would be interested in the confidence intervals associated with the estimate.  Different state points have different levels of noise; having a reliable method for theoretically predicting parameter uncertainty as a function of sample size at different state points  can assist one in designing ``efficient experiments" because one can allocate computational resources in an intelligent manor.  QMLE is a nice tool for achieving all these tasks, however it can fail if it is followed ``verbatim", especially when one uses an approximation of the transition density.  Estimators based on llr methods are appealing in situations where standard QMLE fails because they are applicable to a wider class of problems and the distribution of the llr asymptotically converges to a manageable distribution for certain model classes \cite{lecam86}.

\section{Results and Discussion}
\label{s_results}
%
%


In order to get  MLE (QMLE) parameter estimates in all of the
section to follow, the IMSL Nelder-Mead search algorithm is
employed using the termination criterion of $5\times 10^{-8}$ with
an initial parameter distribution dictated by a uniform
distribution around the slightly biased ($+10\%$) true parameter
values in the cases of known or approximately known models and
from an assumed limit process in the final application.

 \label{results}
\subsection{Maximum Likelihood Estimation Results}
\label{s_resmlebasics}

Tables \ref{tab1} and \ref{tab2}  report the empirically measured mean and standard deviation of the parameter distributions as well as the asymptotic predictions for the standard deviation for situation I A-D and situation I A-C (using data spaced $\delta t=2^{-5} ,2^{-3}$ units respectively).   We see that as $\beta$
increases the estimated mean reversion parameter decreases in magnitude. This makes
sense because the parametric model in equation \ref{eq_cir} assumes a
single free energy minimum; as the magnitude of the sine wave
added to the drift increases, one experiences a situation
where the ``well-depths" associated with the local minima of the glassy potential   increase, retarding the rate
of mean reversion (centered around the global minimum). The exact magnitude of this effect depends
heavily on several factors, some of which are: the magnitude of the noise, the frequency of
the sine wave's perturbation, the amplitude of the sine wave, and  the sampling frequency. The
last two effects are quantified in tables \ref{tab1} and \ref{tab2}.
 Inspection of these tables shows  that the when the true density is used, the
effective model estimated from the data is relatively
independent of the sampling frequency.  Note that as the observation
frequency decreases the quality of
 the expansion (in time)  naturally decreases, but even
 for relatively ``large" times
 between samples A{\"\i}t-Sahalia's expansion
 remains accurate. The quality of the Euler expansion
   is very sensitive to the time between samples and it
  introduces a significant bias even
  when the ``perfect data"
  is presented to the estimator.
    I continue to show the results for the Euler
    estimator despite these shortcomings because
    in section \ref{s_reslaq} the estimator demonstrates
    a  redeeming quality which can be used in conjunction
    with the transition density estimator of A{\"\i}t-Sahalia in  situations
    where the latter fails.
     The tables also show that the sample size is
     large enough to partially realize asymptotic
     results in most cases. 
  Table \ref{tab3} 
  shows similar results;
  the cubic perturbation introduced into the drift results in a higher mean reversion rate.  
  In section \ref{s_reslaq} it is demonstrated that if one presents ``screened" data (those observation pairs that fall within a small neighborhood centered around the estimated $\alpha$ parameter) that  the bias introduced by the nonlinear drift perturbation steadily decreases in magnitude as the neighborhood size decreases which intuitively makes sense given our assumptions.  Unfortunately, this simple procedure creates a more complicated statistical problem in regards to theoretical parameter distribution predictions because  a deterministic approximation of the parameter distribution is much harder to get using this technique.

\begin{table}
\caption{\textbf{Situation I Parameter Distributions} The data used to obtain the parameter distributions was  $N=2000$  sample paths of an SDE sampled over $M=4000$ time intervals evenly spaced at $\delta t=2^{-5}$ taken from the invariant distribution of an Euler path simulation.  The empirical mean and standard deviation  of the parameter distributions are reported as well as the asymptotic predictions of the standard deviation.  For correctly specified models (case A) the asymptotic standard deviation is given by $\mathcal{F}_{OP} $ (calculable by a deterministic integral) and for misspecified models  the standard deviation predicted by $\mathcal{C}$ is reported}

\begin{center} \footnotesize
\begin{tabular}{|l|c|c|c|c|c|c|} \hline
Scenario/ & $<\alpha>$ & $<\kappa>$  &  $<\sigma>$ & $\sigma_{\alpha}^{\mathrm{Asymp}}$ & $\sigma_{\kappa}^{\mathrm{Asymp}}$  &  $\sigma_{\sigma}^{\mathrm{Asymp}}$ \\ \cline{5-7}
\raisebox{1.5ex}[0pt]{Expansion Method} & & &  & $\sigma_{\alpha}^{\mathrm{Emp}}$ & $\sigma_{\kappa}^{\mathrm{Emp}}$  &  $\sigma_{\sigma}^{\mathrm{Emp}}$ \\ \hline \hline

Sit I A&  &   &   & 0.40026&0.26975&0.047621 \\ \cline{5-7} 
\raisebox{1.0ex}[0pt]{True} & \raisebox{2.5ex}[0pt]{20.0052} & \raisebox{2.5ex}[0pt]{4.0428}   & \raisebox{2.5ex}[0pt]{4.0168}&0.40152&0.27501&0.047197\\ \hline 

Sit I A&  &   &   &  0.39938&0.25999&0.047575 \\ \cline{5-7} 
\raisebox{1.0ex}[0pt]{\ait} & \raisebox{2.5ex}[0pt]{20.0454} & \raisebox{2.5ex}[0pt]{3.9628}   & \raisebox{2.5ex}[0pt]{4.0159}&0.40249&0.26681&0.047197\\ \hline 

Sit I A&  &   &   & 0.40000 &0.25298&0.044722 \\ \cline{5-7} 
\raisebox{1.0ex}[0pt]{Euler} & \raisebox{2.5ex}[0pt]{20.0052} & \raisebox{2.5ex}[0pt]{3.7961}   & \raisebox{2.5ex}[0pt]{3.7869}&0.40146&0.24439&0.042119\\ \hline \hline 

Sit I B&  &   &   & 0.40089& 0.27011& 0.047769 \\ \cline{5-7} 
\raisebox{1.0ex}[0pt]{True} & \raisebox{2.5ex}[0pt]{19.9985} & \raisebox{2.5ex}[0pt]{4.0148}   & \raisebox{2.5ex}[0pt]{4.0102}&0.39973&0.26831&0.045601\\ \hline 

Sit I B&  &   &   & 0.3956& 0.24097& 0.047761 \\ \cline{5-7} 
\raisebox{1.0ex}[0pt]{\ait} & \raisebox{2.5ex}[0pt]{20.0374} & \raisebox{2.5ex}[0pt]{3.9362}   & \raisebox{2.5ex}[0pt]{4.0093}&0.40102&0.26296&0.045592\\ \hline 

Sit I B&  &   &   & 0.40097& 0.23932& 0.042992 \\ \cline{5-7} 
\raisebox{1.0ex}[0pt]{Euler} & \raisebox{2.5ex}[0pt]{19.9985} & \raisebox{2.5ex}[0pt]{3.7707}   & \raisebox{2.5ex}[0pt]{3.7822}&0.40052&0.23995&0.041049\\ \hline \hline 

Sit I C&  &   &   & 0.41654& 0.26111& 0.047758 \\ \cline{5-7} 
\raisebox{1.0ex}[0pt]{True} & \raisebox{2.5ex}[0pt]{19.8675} & \raisebox{2.5ex}[0pt]{3.7844}   & \raisebox{2.5ex}[0pt]{3.9277}&0.41787&0.27209&0.047093\\ \hline 

Sit I C&  &   &   &  0.41061& 0.23487& 0.047743 \\ \cline{5-7}
\raisebox{1.0ex}[0pt]{\ait} & \raisebox{2.5ex}[0pt]{19.9016} & \raisebox{2.5ex}[0pt]{3.7085}   & \raisebox{2.5ex}[0pt]{3.9267}&0.4196&0.27554&0.047109\\ \hline 

Sit I C&  &   &   & 0.42007& 0.2335& 0.043175 \\ \cline{5-7}
\raisebox{1.0ex}[0pt]{Euler} & \raisebox{2.5ex}[0pt]{19.8674} & \raisebox{2.5ex}[0pt]{3.5374}   & \raisebox{2.5ex}[0pt]{3.7195}&0.41835&0.24537&0.041998\\ \hline \hline 

Sit I D&  &   &   &   0.49515& 0.22916& 0.047533 \\ \cline{5-7}
\raisebox{1.0ex}[0pt]{True} & \raisebox{2.5ex}[0pt]{19.3406} & \raisebox{2.5ex}[0pt]{2.9368}   & \raisebox{2.5ex}[0pt]{3.581}&0.48422&0.27015&0.058569\\ \hline 

Sit I D&  &   &   &  0.48505& 0.21071& 0.047373 \\ \cline{5-7}
\raisebox{1.0ex}[0pt]{\ait} & \raisebox{2.5ex}[0pt]{19.3614} & \raisebox{2.5ex}[0pt]{2.8953}   & \raisebox{2.5ex}[0pt]{3.5787}&0.48679&0.26281&0.050485\\ \hline 

Sit I D&  &   &   & 0.51899& 0.21057& 0.0439 \\ \cline{5-7} 
\raisebox{1.0ex}[0pt]{Euler} & \raisebox{2.5ex}[0pt]{19.3387} & \raisebox{2.5ex}[0pt]{2.6775}   & \raisebox{2.5ex}[0pt]{3.4387}&0.48306&0.24238&0.044822\\ \hline

\end{tabular}
\end{center}
\label{tab1}
\end{table}

\begin{table}
\caption{\textbf{Situation I Parameter Distributions} Same information as table \ref{tab1} except the time intervals are evenly spaced at $\delta t=2^{-3}$.  Note how the parameter distribution quality  degrades (as compared to table \ref{tab1}), this is due in part to the failure of the transition density expansion (as evident from situation I A).  Another thing to note is that when the true CIR density is used, the parameters estimated are relatively independent of the sampling frequency (which is a very desirable quality).  This is not the case for the two transition density expansions, however the magnitude of the discrepancy between this table and the previous is much smaller for the \ait expansion. }

\begin{center} \footnotesize
\begin{tabular}{|l|c|c|c|c|c|c|} \hline
Scenario/ & $<\alpha>$ & $<\kappa>$  &  $<\sigma>$ & $\sigma_{\alpha}^{\mathrm{Asymp}}$ & $\sigma_{\kappa}^{\mathrm{Asymp}}$  &  $\sigma_{\sigma}^{\mathrm{Asymp}}$ \\ \cline{5-7}
\raisebox{1.0ex}[0pt]{Expansion Method} & & &  & $\sigma_{\alpha}^{\mathrm{Emp}}$ & $\sigma_{\kappa}^{\mathrm{Emp}}$  &  $\sigma_{\sigma}^{\mathrm{Emp}}$ \\ \hline \hline

Sit I A&  &   &   & 0.20206&0.16652&0.056847 \\ \cline{5-7} 
\raisebox{1.0ex}[0pt]{True} & \raisebox{2.5ex}[0pt]{19.9996} & \raisebox{2.5ex}[0pt]{4.0296}   & \raisebox{2.5ex}[0pt]{4.019}&0.20085&0.16673&0.057301\\ \hline 

Sit I A&  &   &   & 0.18208&0.10863&0.055179 \\ \cline{5-7} 
\raisebox{1.0ex}[0pt]{\ait} & \raisebox{2.5ex}[0pt]{20.1162} & \raisebox{2.5ex}[0pt]{3.6394}   & \raisebox{2.5ex}[0pt]{3.9971}&0.22812&0.23505&0.06125\\ \hline 

Sit I A&  &   &   & 0.20088&0.12692&0.045736 \\ \cline{5-7} 
\raisebox{1.0ex}[0pt]{Euler} & \raisebox{2.5ex}[0pt]{19.9996} & \raisebox{2.5ex}[0pt]{3.1652}   & \raisebox{2.5ex}[0pt]{3.2323}&0.20088&0.10329&0.038508\\ \hline \hline 

Sit I B&  &   &   & 0.7049& 0.4502& 0.03758 \\ \cline{5-7} 
\raisebox{1.0ex}[0pt]{True} & \raisebox{2.5ex}[0pt]{19.9915} & \raisebox{2.5ex}[0pt]{4.0095}   & \raisebox{2.5ex}[0pt]{4.0123}&0.20131&0.1659&0.056841\\ \hline 

Sit I B&  &   &   & 0.76493& 0.41996& 0.037335 \\ \cline{5-7} 
\raisebox{1.0ex}[0pt]{\ait} & \raisebox{2.5ex}[0pt]{20.118} & \raisebox{2.5ex}[0pt]{3.6523}   & \raisebox{2.5ex}[0pt]{3.9941}&0.22589&0.21442&0.059281\\ \hline 

Sit I B&  &   &   & 0.30418& 0.26104& 0.049705 \\ \cline{5-7} 
\raisebox{1.0ex}[0pt]{Euler} & \raisebox{2.5ex}[0pt]{19.9915} & \raisebox{2.5ex}[0pt]{3.1517}   & \raisebox{2.5ex}[0pt]{3.2306}&0.20138&0.10267&0.038656\\ \hline \hline 

Sit I C&  &   &   & 0.32825& 0.29782& 0.048524 \\ \cline{5-7} 
\raisebox{1.0ex}[0pt]{True} & \raisebox{2.5ex}[0pt]{19.8718} & \raisebox{2.5ex}[0pt]{3.7377}   & \raisebox{2.5ex}[0pt]{3.9093}&0.20573&0.16222&0.055943\\ \hline 

Sit I C&  &   &   & 0.33702& 0.26354& 0.041954 \\ \cline{5-7} 
\raisebox{1.0ex}[0pt]{\ait} & \raisebox{2.5ex}[0pt]{19.9835} & \raisebox{2.5ex}[0pt]{3.4578}   & \raisebox{2.5ex}[0pt]{3.8975}&0.22621&0.21599&0.05993\\ \hline 

Sit I C&  &   &   & 0.38195& 0.24293& 0.04877\\ \cline{5-7} 
\raisebox{1.0ex}[0pt]{Euler} & \raisebox{2.5ex}[0pt]{19.8717} & \raisebox{2.5ex}[0pt]{2.9659}   & \raisebox{2.5ex}[0pt]{3.1967}&0.20574&0.10375&0.03815\\ \hline

\end{tabular}
\end{center}
\label{tab2}
\end{table}

\begin{table}
\caption{\textbf{Situation II Parameter Distributions} 
Same information as table \ref{tab1} except $M=4500$.} 
\begin{center} \footnotesize
\begin{tabular}{|l|c|c|c|c|c|c|} \hline
Scenario/ & $<\alpha>$ & $<\kappa>$  &  $<\sigma>$ & $\sigma_{\alpha}^{\mathrm{Asymp}}$ & $\sigma_{\kappa}^{\mathrm{Asymp}}$  &  $\sigma_{\sigma}^{\mathrm{Asymp}}$ \\ \cline{5-7}
\raisebox{1.5ex}[0pt]{Expansion Method} & & &  & $\sigma_{\alpha}^{\mathrm{Emp}}$ & $\sigma_{\kappa}^{\mathrm{Emp}}$  &  $\sigma_{\sigma}^{\mathrm{Emp}}$ \\ \hline \hline

Sit II A&  &   &   & 0.37737&0.25432&0.044898 \\ \cline{5-7} 
\raisebox{1.0ex}[0pt]{True} & \raisebox{2.5ex}[0pt]{19.9877} & \raisebox{2.5ex}[0pt]{4.0512}   & \raisebox{2.5ex}[0pt]{4.0169}&0.38691&0.26327&0.04653\\ \hline 

Sit II A&  &   &   & 0.37654&0.24512&0.044854 \\ \cline{5-7} 
\raisebox{1.0ex}[0pt]{\ait} & \raisebox{2.5ex}[0pt]{20.0272} & \raisebox{2.5ex}[0pt]{3.9695}   & \raisebox{2.5ex}[0pt]{4.0159}&0.38803&0.25535&0.046481\\ \hline 

Sit II A&  &   &   & 0.37712&0.23852&0.042164 \\ \cline{5-7} 
\raisebox{1.0ex}[0pt]{Euler} & \raisebox{2.5ex}[0pt]{19.988} & \raisebox{2.5ex}[0pt]{3.8045}   & \raisebox{2.5ex}[0pt]{3.7862}&0.38732&0.23417&0.041944\\ \hline \hline 

Sit II B&  &   &   & 0.30542& 0.28151& 0.045772 \\ \cline{5-7} 
\raisebox{1.0ex}[0pt]{True} & \raisebox{2.5ex}[0pt]{19.826} & \raisebox{2.5ex}[0pt]{4.9256}   & \raisebox{2.5ex}[0pt]{4.018}&0.31247&0.26417&0.047077\\ \hline 

Sit II B&  &   &   & 0.29819& 0.25567& 0.045828 \\ \cline{5-7} 
\raisebox{1.0ex}[0pt]{\ait} & \raisebox{2.5ex}[0pt]{19.8544} & \raisebox{2.5ex}[0pt]{4.8679}   & \raisebox{2.5ex}[0pt]{4.0186}&0.30856&0.25292&0.047127\\ \hline 

Sit II B&  &   &   & 0.30545& 0.24273& 0.039965 \\ \cline{5-7} 
\raisebox{1.0ex}[0pt]{Euler} & \raisebox{2.5ex}[0pt]{19.826} & \raisebox{2.5ex}[0pt]{4.5645}   & \raisebox{2.5ex}[0pt]{3.7382}&0.31238&0.22756&0.040756\\ \hline

\end{tabular}
\end{center}
\label{tab3}
\end{table}

\subsection{Optimal Binary Alternative Hypothesis Testing Results}
\label{s_resnp}
\begin{figure}
\includegraphics[angle=0,scale=.65]{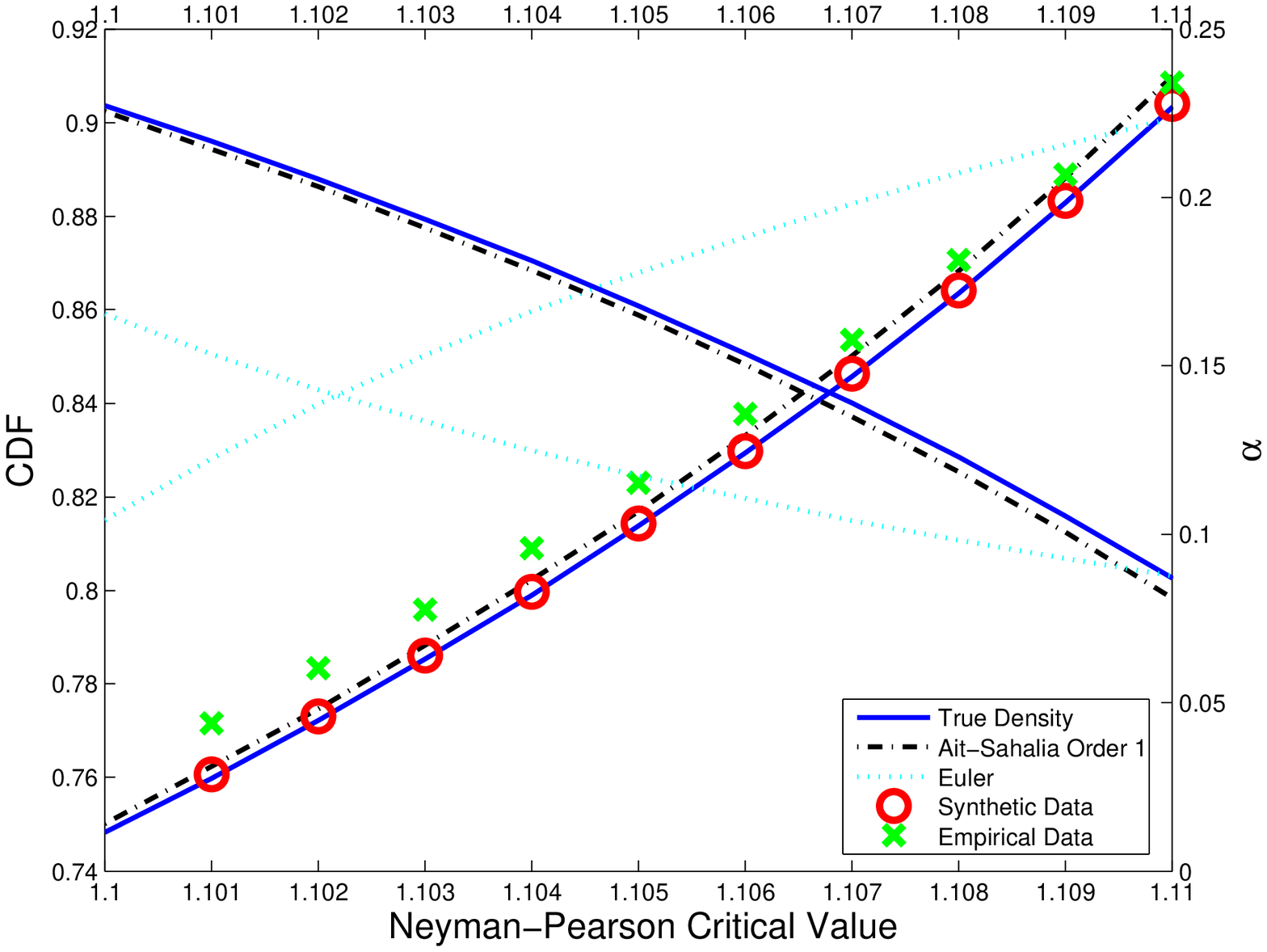}
\caption{Neyman-Pearson Results \small{
\emph{The Neyman-Pearson test carried out on
situation II B data using the QMLE parameter estimate as the
alternative and the exact (known) parameters associated with situation II A as the null.  The
curves represent the deterministic calculation of the type I error probability as a function of the critical value (right axis) and
the theoretical CDF of the likelihood ratio obtained by using the
various transition density expansions assuming ergodic sampling of the
invariant distribution (left axis).  The ``x's" correspond to using the actual
situation II B data (nonlinear potential) and the ``o's"
correspond to the distribution of the likelihood ratio obtained using the
QMLE parameters estimated  to simulate sample paths of the
parametric model proposed (plugging in the average of the QMLE parameter distribution) and then using the exact  CIR density to evaluate the likelihood ratio. }} 
}

\label{f_npcrit}
\end{figure}

In figure \ref{f_npcrit}, the top/right axis
 corresponds to
the Neyman-Pearson critical value versus
the theoretical type I error probability
for all three transition density expansions
using the null as the true (known) parameters of situation II A
and the alternative as the parameters obtained using situation II B data with QMLE (using the exact CIR density).
 The left/bottom axis shows the analytically calculated cumulative
  distribution of rejecting the null under the alternative using the
   three different transition densities (plugging in the same alternative parameters estimated by QMLE)  as well as the
   empirically measured distribution of the likelihood ratio
   (see caption for additional details).
   We see that the sample size is large enough
   to realize agreement between the
   measured and limit distributions.
    The expansion  of A{\"\i}t-Sahalia provides a
    very good approximation of this distribution,
    whereas the distribution predicted by the Euler approximation
    deviates substantially.

\subsection{Goodness-of-fit and Model Misspecification Results}
\begin{figure}
\includegraphics[angle=0,scale=.65]{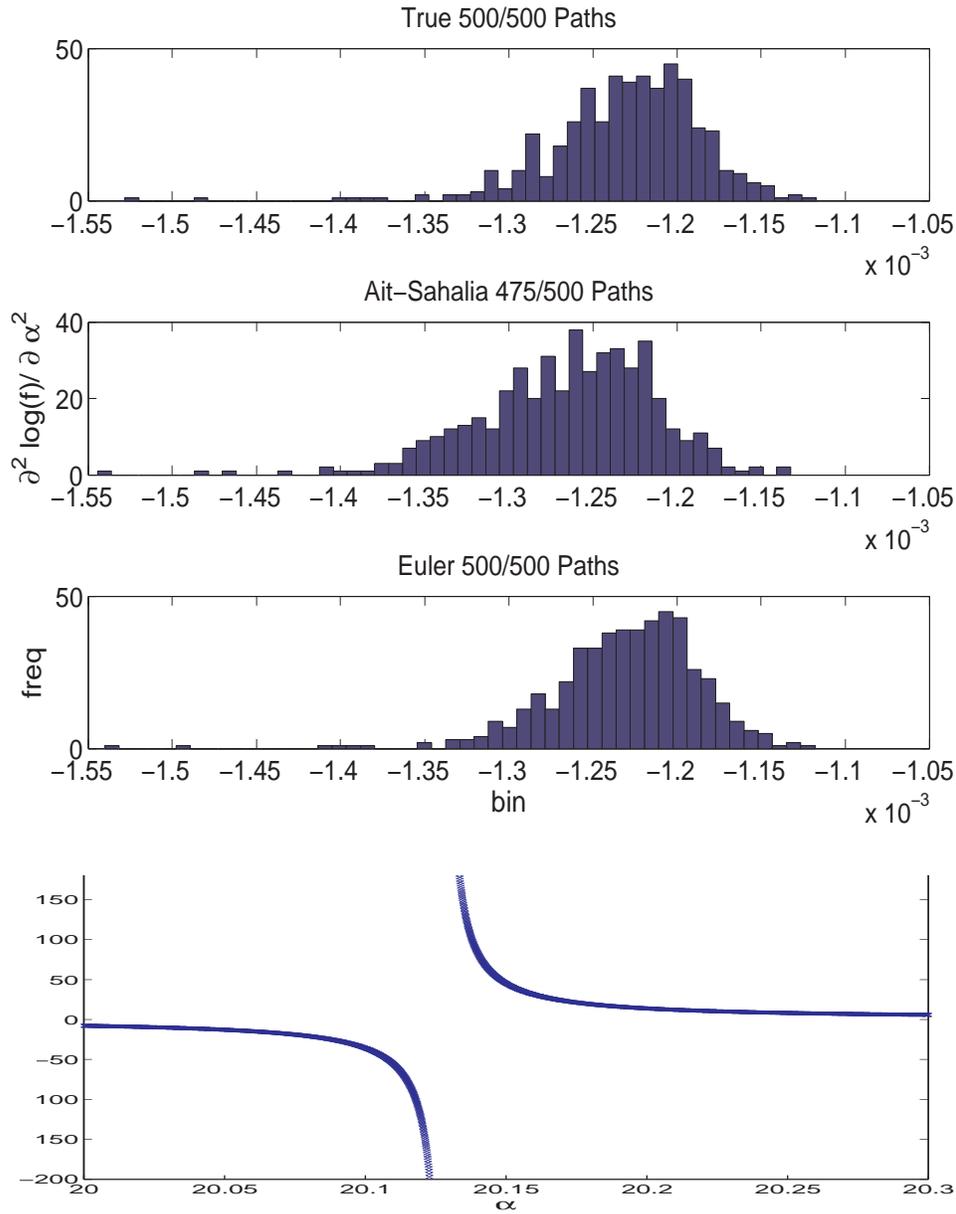}
\caption{Histogram of Diagonal Component of $\mathcal{F}_{Hessian}$ Corresponding to $\alpha$ for the Three Transition Density Expansions \small{
\emph{  The histograms all have slightly different shapes indicating that the curvatures of the three transition densities differ (data taken from situation I C with $\delta t=2^{-5}$).  In the second plot 25 observations were thrown out because there were outliers caused by spurious singularities in the transition density expansion. The bottom panel gives one demonstration of how the \ait expansion can introduce spurious singularities into the log-likelihood function.  The data shown is the second derivative of transition density expansion with respect to $\alpha$ as a function of $\alpha$ for a particular observation (see text for discussion).  The fat tails of the log likelihood function  make a simple screening  of outliers (caused by spurious singularities) very difficult if one does not have the true transition density to reference.} 
}}

\label{f_singdist}
\end{figure}

 Here it is demonstrated how the various transition densities
perform when one tries to use them to evaluate the derivatives needed to estimate some goodness-of-fit statistics
shown in \cite{white}.
Before proceeding,  a mildly problematic aspect of the
expansion of A{\"\i}t-Sahalia is pointed out.  To illustrate the problem, 
the histograms of $\frac{\partial^2\mathcal{L_\theta}}{\partial \alpha ^2}$ are plotted for 500 sample paths using the three
expansions.  The particular histogram shown  using the
A{\"\i}t-Sahalia expansion had 25 observations that were
disregarded because of
unusually high values in the calculated quantities.   The expansion of \ait is very accurate, but
its functional derivatives can unfortunately introduce spurious singularities into the transition density approximation.  To
show a specific example of this, take the logarithm of the order
one A{\"\i}t-Sahalia CIR expansion \footnote{Mathematica code
available from $\mathrm{http://www.princeton.edu/ \widetilde{ }yacine/research.htm}$} and calculate the
second derivative with respect to $\alpha$ and plug in the
observation pair (obtained via an Euler path simulation)
$[x_n,x_{n-1}]=[3.1826960,3.305275]$ and set $\sigma=4,\kappa=4,\delta t=2^{-5}$. The bottom panel in figure
\ref{f_singdist}  plots the resulting function of $\alpha$.   For
the particular model and parameter values shown, 
this type of situation is only  encountered when calculating functional derivatives.
In the transition density associated with the model given in equation \ref{eq_myaffine},  sample
values where singularities are hit when evaluating the pure log
likelihood function are encountered.  A standard method for dealing with this is
to apply a one-step MLE estimator \footnote{The basic ideas of
these procedures is to use a simple restricted parameter set that
is ``rich" in the parameter space \cite{semipar}.  An example of
this would be to take a discrete mesh of parameter space and
optimize the log likelihood over this finite set.  The main
idea is to keep the parameter values away from singularities
associated with the finite sample log likelihood.  Refer to the literature on asymptotically centering estimators in \cite{lecam00,lecam60} for another example of a one-step method.  }.
 A minor modification of Le Cam's
method shown in section \ref{s_laq} provides one possible construction of a
one-step MLE estimator.
   Many one-step methods require
 a parameter guess that is within a ``reasonable neighborhood" of the
true parameter value (the exact size of the neighborhood can be chosen using
an approximation of the Fisher information matrix).  Table \ref{tabonestep}
illustrates that if one starts with a slightly perturbed guess of the true parameter that Le Cam's method can get fairly close to the true parameter vector if the transition density approximation is very accurate (the llr expansion was obtained from situation II A data and the one-step used is outlined in Le Cam \cite{lecam00} chapter 6). The asymptotic likelihood expansions are  valuable  in parametric estimation; unfortunately this application requires an extremely accurate transition density approximation \footnote{ Instead of applying one-step methods (a full discussion would slightly overload this paper and distract from the simpler points), in the simple
parametric models studied I remedy the singularity issue by  using the Euler approximation to
determine when a singularity is hit. It is possible to distinguish between spurious
singularities from log likelihood function singularities in the CIR case because
the true transition density is known. In all of the CIR applications the
singularities hit were in fact spurious.  I simply threw out
sample paths where the absolute value of the logarithm of the
transition density of the A{\"\i}t-Sahalia expansion differed from
that of the Euler approximation by a factor of three anywhere along the discretely observed path (this occurred less than $4\%$ in the CIR studies, in the SSA studies this number increased to roughly $25\%$).  
\label{throwout}
}. Throughout this paper the first order (in time) \ait expansions have been used; only in table \ref{tabonestep} do I report results from a higher order expansion (order four). Recall in section \ref{f_npcrit} that the order one \ait expansion provided an accurate approximation of the likelihood ratio for a simple hypothesis test.  The test statistic generated for the Neyman-Pearson test only required the ratio of one observation pair at two nearby parameter vectors.  The transition density expansion does introduce a systematic bias into the approximation, and the nature the bias does exhibit a smooth dependence on the parameter values making the ratio of two nearby probability densities also exhibit a significant bias and
these small errors accumulate as the time series
length grows (see equation \ref{eq_loglikelihoodfuncdef}) complicating matters.   Some techniques and analysis have already been developed for approximating the LAQ expansion of random variables in the case where the density is not known explicitly \cite{jeganathan}; most techniques developed require one to empirically measure the transition density.  For our applications, the empirical distribution techniques are mildly  inconvenient (from a computational standpoint) because one needs to determine an empirical density approximation for each observation pair.

\begin{table}
\caption{\textbf{Le Cam LAQ One-Step Test} The data used was situation II A with $\delta t=2^{-5}$.  The LAQ motivated one-step expansion was carried out at the point $\theta=(20.1,4.5,4.1)$, consult Le Cam \cite{lecam00} chapter 6 for details.  The one-step estimate of the parameter results from adding the quantity below to  $\theta$.  This type of procedure is necessary when standard MLE misbehaves.  The table shows that the LAQ update gets one close to the true parameter when an extremely accurate approximation of the  transition density is in hand.  The \ait order one and Euler updates are not as good, but still usable.  The major problem with these estimators is that the contiguity condition becomes difficult to verify (a condition that needs to be met before the LAQ expansion can be used with confidence).  The last two columns display -2 times the mean and the variance of the llr measured by evaluating it at $\hat{\theta}$ and $\hat{\theta}+h_M$  where $h_M=(1.2,0,0.12)$.  In the infinite sample limit these two quantities will be identical (assuming $h_M$ is continually scaled properly). For the exact and order four \ait expansion one observes close results.  For the other two expansions this is not the case (see text for further discussion).}

\begin{center} \footnotesize
\begin{tabular}{|l|c|c|c|c|c|} \hline
Expansion & $\Delta \alpha$ & $\Delta \kappa$  &  $\Delta \gamma$ & $-2\mu_{\Lambda_{\theta^{\mathrm{true}}}}$ & $\sigma_{\Lambda_{\theta^{\mathrm{true}}}}$ \\ \hline
 
CIR Exact  &-0.9507 &   -0.4791 &   -0.0931 & 12.9402 & 13.8679 \\ \hline
\ait Order 4  & -0.9507 &   -0.4791 &   -0.0931 & 12.9405  & 13.8679  \\ \hline
\ait Order 1  & -1.0065 &  -0.7499 &  -0.1208 & 13.6013 & 39.0793  \\ \hline
Euler  & -0.9621 &  -0.7344 &   -0.3156 &  41.8144 & 16.0470  \\ \hline

\end{tabular}
\end{center}
\label{tabonestep}
\end{table}

\begin{figure}
\includegraphics[angle=0,scale=.65]{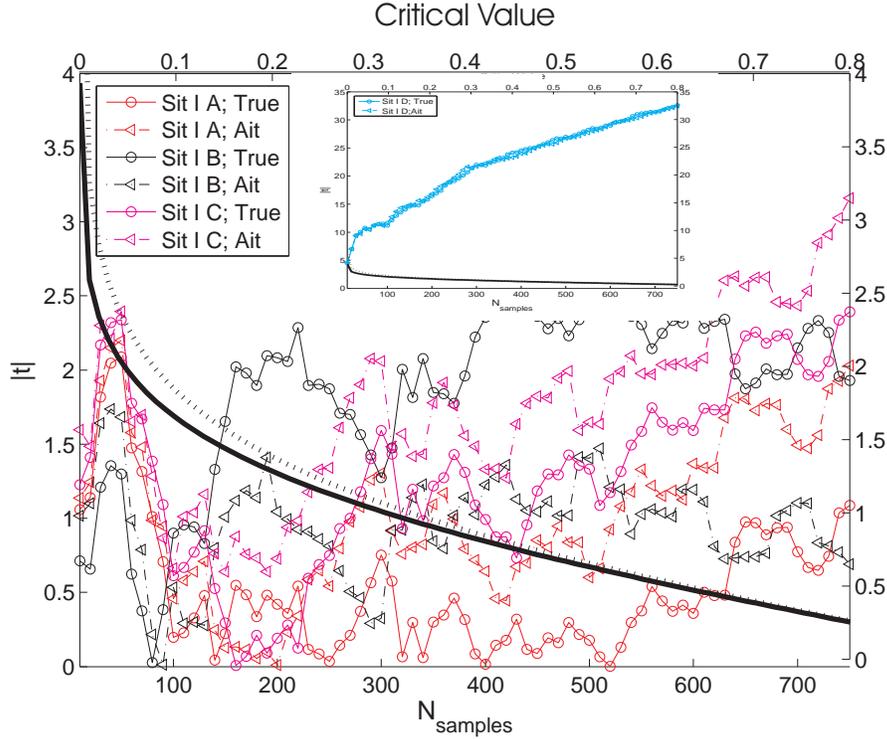}
\caption{Model Misspecification t-test Results \small{
\emph{The left/bottom axis plots the t-statistic obtained by summing the components of $\mathcal{F}_{OP}$ and $\mathcal{F}_{Hessian}$ for various sample sizes (time series length is fixed, different $N_{samples}$ values correspond to using different sample path numbers to create the random variables required to generate the sample mean and standard deviation needed for the t-test ).  The right/top axis is to be used with the lines without symbols; the top axis is the critical value (cv) of the t-test (lowest $|t|$ needed to reject the null for a given $\alpha$). Samples were drawn at random, but once a $T$ r.v. was drawn it contributed to the cumulative average plotted. The solid line corresponds to the theoretical cv of a sample size =375 and the dashed line corresponds to a sample size =10. The inset plots the situation I D test which can be rejected quickly with high confidence.   } 
}}

\label{f_twhite}
\end{figure}

Now let us return to the goodness-of-fit issue. First, the condition $\mathcal{F}_{Hessian}=- \mathcal{F}_{OP}$
it tested.  If the
condition holds $T:= \sum\limits_{i,j=1}^{k}\mathcal{F}_{Hessian}^{ij}+\mathcal{F}_{OP}^{ij}$ should be a mean zero random variable (the superscripts denote  matrix components and $k$ is the number of parameters estimated).
If it is assumed that the classical central limit theorem (CLT) holds and that the sample sizes used are large enough to appeal
to the CLT and approximate the sample mean by a
normal distribution of unknown variance, then  the
classical t-test can be employed \cite{bickel} (more
sophisticated tests are proposed in \cite{white}).
 Figure \ref{f_twhite} plots
the t-test results for testing if ${F}_{Hessian}=- \mathcal{F}_{OP}$ for various sample sizes (the A{\"\i}t-Sahalia
data is screened by the technique  mentioned in footnote \ref{throwout}).
 The figure shows that as the number of paths grows (the equality tested is valid in the infinite time series length limit; here the time series length is held fixed), that it
becomes easier to detect discrepancies between the data and the assumed
model.  Even when the true transition density is used with the correct model, one observes that as sample size increases (in paths)  that it becomes easier to reject the null. This is because the equality $\mathcal{F}_{Hessian}=- \mathcal{F}_{OP}$ is an asymptotic result.  A finite size time series can \emph{never} fully realize asymptotic results; as more paths are analyzed the departure from the limit becomes easier to detect.
  One should also note that as the number of paths increases  it also becomes easier to detect the errors in the transition density expansion.  Before proceeding to check for model misspecication in the presence of an approximated transition density,  one should determine if  asymptotic results  are valid for ``perfect data" using  the same time series length, then proceed to check the accuracy of the expansion by some MC tests (like the ones presented in the previous section). 

\begin{figure}
\includegraphics[angle=0,scale=.65]{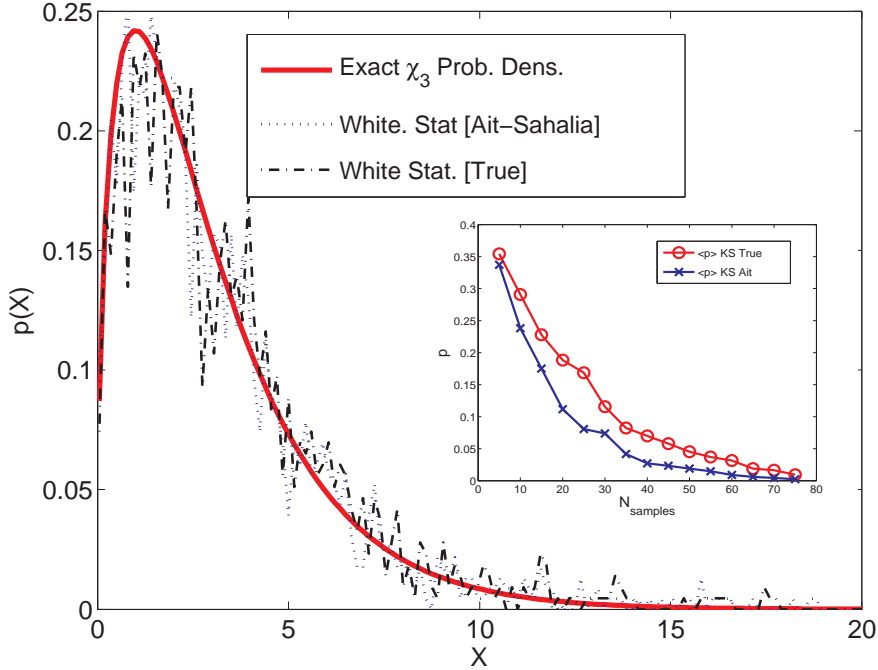}
\caption{Goodness-of-Fit with Kolmogorov-Smirnov Test \small{
\emph{ The $\chi^2(3)$ probability density is plotted along with the empirical distribution of the test-statistic proposed in \cite{white} for situation I C data.  The inset shows a plot of the p-value (minimum $\alpha$ needed to reject the null given the data) obtained using the one-sample Kolmogorov-Smirnov test versus the number of paths used to create the test statistic.  If the full set of samples were used, the test would have no problem in rejecting the null.  This is due to the fact that asymptotic results are never truly realized with finite sample sizes.  The plot illustrates that the test statistic generated by the \ait expansion does faithfully capture the average curvature of the log likelihood function in a misspecified model.  } 
}}

\label{f_chidistwhite}
\end{figure}

For  extreme cases (situation I C and D) the t-test safely rejects the
null for  small  sizes (see figure
\ref{f_twhite} and the inset).  This indicates that the data is well outside the
particular parametric model class under study, but one can still
test if some asymptotic QMLE results hold.  Figure \ref{f_chidistwhite}
plots the  exact probability density  of the $\chi^2(3)$ random variable  as well
a test statistic given in \cite{white} (which will be denoted by $\mathcal{HW}$ \footnote{ $\mathcal{HW}:=Mg(\theta)\Big[\nabla g (\theta) \mathcal{C}_{(\theta)} \nabla g(\theta)^{T}\Big]^{-1}g(\theta)^T$   where $g(\cdot)$ is the gradient (written as a row vector) of the log likelihood function.  If $\theta \in \mathbb{R}^k$ then under conditions stated in \cite{white} this random variable converges in distribution to a $\chi^2(k)$  random variable.  This statistic was originally proposed by Halbert White \cite{white}.} )
  We see that both the approximation and the
exact transition density appear close to the predicted limit
distribution indicating that the asymptotic results approximately
hold for the sample sizes used.  I quantitatively compare how
well the empirical distributions match the limit distribution by
using the Kolmogorov-Smirnov (KS) test with various sample sizes
in the inset of figure \ref{f_chidistwhite}  \footnote{In this particular application, we
are still significantly away from the limit distribution. The
sample sizes used for estimation are large enough that a test as
powerful as the KS would  reject equality of the two distributions
with very high certainty if one used all of the data at hand.  For this reason I only present
portions of the data to the KS test. In many interesting atomistic simulation studies, one can only afford to generate a couple hundred sample paths making this type of goodness-of-fit test useful in practice.  An alternative would be to
partition the CDF into bins and use the $\chi^2$ square goodness-of-fit test \cite{lehmann}.    }.  The average p-value is obtained by applying the one sample KS test \cite{bickel} using 500 draws (with replacement) of size $N_{samples}$ from the pool of $\mathcal{HW}$ random variables associated with the $N_{samples}$ paths  using the  $\chi^2(3)$ density as the null.  The result of this procedure is shown in the inset of figure \ref{f_chidistwhite}.  We see that the test
statistic created using the A{\"\i}t-Sahalia expansion is rejected
before that associated with the true density indicating the
possibility that the errors in the expansion cause a \emph{
mildly} inflated rejection rate in situation I C.

\subsection{Le Cam's Method and Likelihood Ratio Expansions Results}
\label{s_reslaq}
Table \ref{tabcirhit} presents parameter estimates of
the screened data
(the neighborhood size used is given in the captions) from situation II B.
 We see that as the neighbrohood size decreases
 the estimator mean moves closer to that of the
 ``uncontaminated" model but the variance of the
 estimator increases.
   I use Le Cam's LAQ expansion with the
   \ait transition density
   expansions in order to quantify the uncertainty in the measurement.  One should observe that the LAQ matrix is closely related to 
   the variance of the parameter distributions (for a more precise description of the connection consult \cite{jeganathan,basawa}).
   The agreement between the estimated and measured parameter variance is not as sharp as
   it was in the case of a stationary distribution, but there one had the convenience of evaluating a deterministic integral (in situation A).  Furthermore the asymptotic results are harder to realize because some observation pairs are not used to obtain parameter estimates (due to the screening).
If one has a situation where the neighborhood size is small enough to give ``meaningful" QMLE parameter estimates and the number of observation pairs is large enough to appeal to asymptotic methods, then
     the
    LAQ screening method can be used to
    define a cost function which
    can be used to intelligently choose the screening neighborhood size.
\begin{table}
\caption{\textbf{Situation II B Parameter Distributions with LAQ Screening} The data used to obtain the parameter distributions was the same as that used in table \ref{tab3} except here only results of using the \ait expansion are presented using screened data.  The base point $(X_o)$ of all of the local SDE models shown below is 20 which corresponds to the true $\alpha$ parameter and the interval used to filter the data is given in the first left column.  The empirical  mean and standard deviation of the parameters  are given along with the LAQ prediction of the parameter uncertainty. }

\begin{center} \footnotesize
\begin{tabular}{|l|c|c|c|c|c|c|} \hline
 &  & $ $  &  $ $ & $\sigma_{\alpha}^{\mathrm{Asymp}}$ & $\sigma_{\alpha}^{\mathrm{Asymp}}$  &  $\sigma_{\kappa}^{\mathrm{Asymp}}$ \\ \cline{5-7}
\raisebox{2.5ex}[0pt]{Neighborhood Size} & \raisebox{2.5ex}[0pt]{$<\alpha>$} & \raisebox{2.5ex}[0pt]{$<\kappa>$} & \raisebox{2.5ex}[0pt]{$<\sigma>$}  & $\sigma_{\alpha}^{\mathrm{Emp}}$ & $\sigma_{\kappa}^{\mathrm{Emp}}$  &  $\sigma_{\sigma}^{\mathrm{Emp}}$ \\ \hline \hline

&  &   &   & 0.577& 1.079& 0.094 \\ \cline{5-7} 
\raisebox{2.5ex}[0pt]{$(16,24)$} & \raisebox{1.5ex}[0pt]{19.993} & \raisebox{1.5ex}[0pt]{4.080} & \raisebox{1.5ex}[0pt]{4.006} & 0.560& 1.008 & 0.085 \\ \hline

&  &   &   & 0.502& 0.780& 0.078 \\ \cline{5-7} 
\raisebox{2.5ex}[0pt]{$(15,25)$} & \raisebox{1.5ex}[0pt]{20.047} & \raisebox{1.5ex}[0pt]{4.228} & \raisebox{1.5ex}[0pt]{4.003} & 0.530 & 0.802 & 0.078 \\ \hline

&  &   &   & 0.450& 0.629& 0.068 \\ \cline{5-7} 
\raisebox{2.5ex}[0pt]{$(14,26)$} & \raisebox{1.5ex}[0pt]{20.021} & 
\raisebox{1.5ex}[0pt]{4.365} & \raisebox{1.5ex}[0pt]{4.006} & 0.474 &0.627 & 0.066 \\ \hline


\end{tabular}
\end{center}
\label{tabcirhit}
\end{table}

\subsection{Local Polynomial Diffusion Models  Results}
\label{s_reslocal}

\begin{figure}
\includegraphics[angle=0,scale=.85]{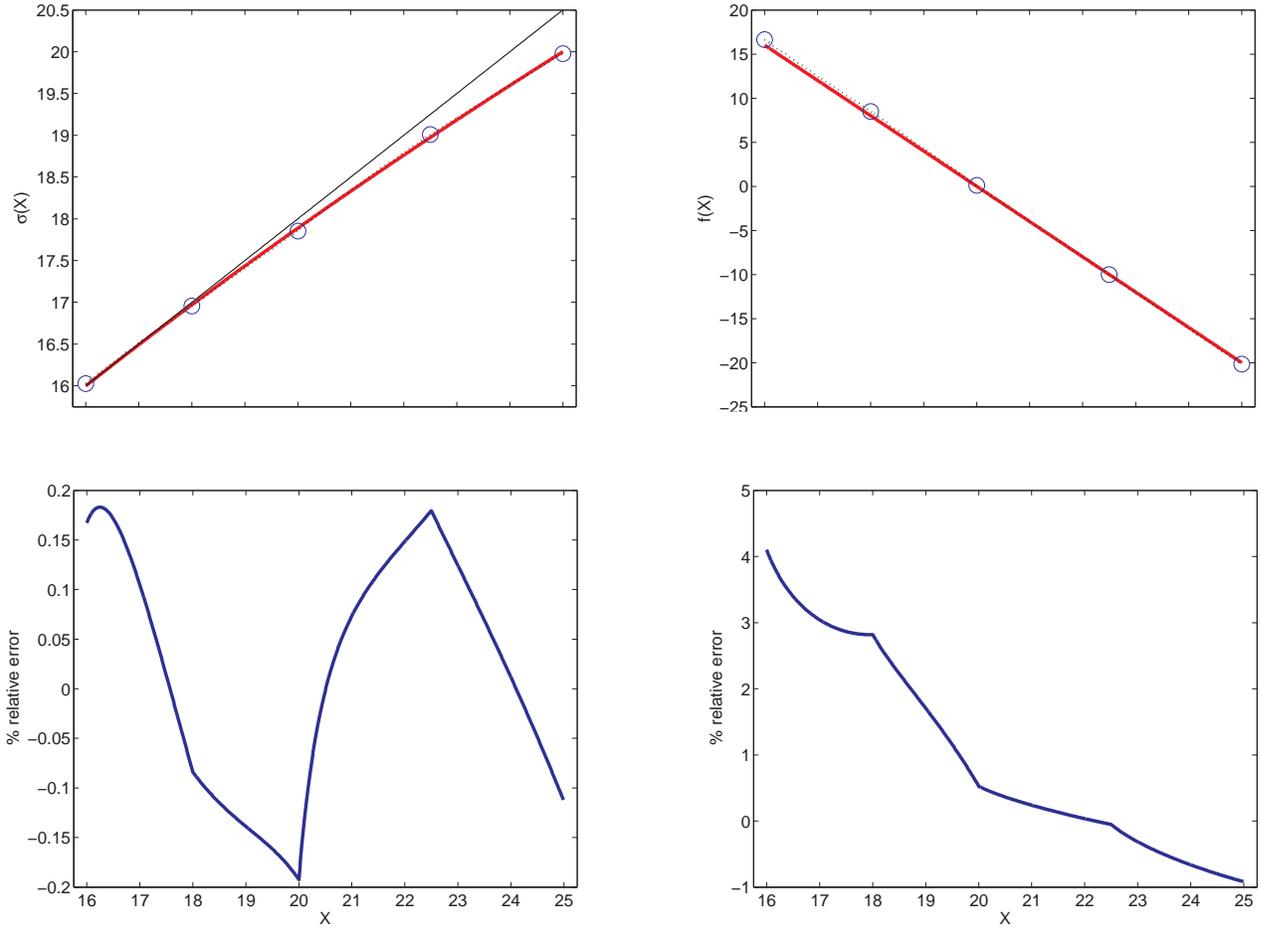}
\caption{Piecewise Polynomial Approximation of CIR Model \small{
\emph{The top  panels plot the diffusion (left) and drift (right)  function obtained by the interpolation procedure given in section \ref{s_reslocal}.  The top left panel plots the first order Taylor expansion (evaluated at $X_o=16 $) of the diffusion coefficient from the known function to show how much the true function deviates from linearity and how well the procedure detects this change.  The bottom figures plot the corresponding relative errors (using the exact known SDE coefficient functions).  See footnote \ref{cireulexplanation} for a discussion on systematic versus random errors. 
}}}
\label{f_CIRpieces}
\end{figure}

In the first application, the pp SDE method is applied to data generated by
the unperturbed CIR model ($\gamma=\beta=0$);
I pretend that the functional form of
the drift or diffusion coefficient of the data generating process is unknown and instead use the model in equation \ref{eq_myaffine}.
Five arbitrary state points denoted  
 by $\{ E_i \}_{i=1,5}$
(shown  as circles in figure \ref{f_CIRpieces}) are chosen
and the parameters of the affine SDE are obtained there.  The
LAQ expansion is used in order to obtain parameter
uncertainty estimates  and
the results are compared to the
 empirical  parameter
distribution measured by carrying out QMLE on the screened data in table \ref{ppSDE} .

\begin{table}
\caption{\textbf{Piecewise Polynomial SDE Parameter Distributions of Situation II B} The data used to obtain the parameter distributions was  $N=2000$  sample paths of an SDE sampled over $M=4000$ time intervals evenly spaced at $\delta t=2^{-5}$ using expansion point $X_o$ with the neighborhood size given in the second column.  The empirical mean and standard deviation  of the parameter distributions are reported as well as the LAQ predictions of the standard deviation.} 

\begin{center} \footnotesize
\begin{tabular}{|c|c|c|c|c|c|c|c|c|c|} \hline

 &   & &  &   &  & $\sigma_{a}^{\mathrm{Asymp}}$ & $\sigma_{b}^{\mathrm{Asymp}}$  &  $\sigma_{c}^{\mathrm{Asymp}}$ & $\sigma_{d}^{\mathrm{Asymp}}$ \\ \cline{7-10}
\raisebox{2.5ex}[0pt]{Xo} & \raisebox{2.5ex}[0pt]{Neighborhood} & \raisebox{2.5ex}[0pt]{$<a>$} & \raisebox{2.5ex}[0pt]{$<b>$} & \raisebox{2.5ex}[0pt]{$<c>$}  & \raisebox{2.5ex}[0pt]{$<d>$} & $\sigma_{a}^{\mathrm{Emp}}$ & $\sigma_{b}^{\mathrm{Emp}}$  &  $\sigma_{c}^{\mathrm{Emp}}$ & $\sigma_{d}^{\mathrm{Emp}}$  \\ \hline \hline

  &  &  &   &   & &  1.939 & 0.696 & 0.295 & 0.067 \\ \cline{7-10} 
\raisebox{2.5ex}[0pt]{16} & \raisebox{2.5ex}[0pt]{$(11,22)$} & \raisebox{2.5ex}[0pt]{15.909} & \raisebox{2.5ex}[0pt]{-4.227} & \raisebox{2.5ex}[0pt]{15.600} & \raisebox{2.5ex}[0pt]{0.534} & 1.999&0.713&0.278& 0.061\\ \hline \hline

  &  &  &   &   & &  1.898 & 0.6344 & 0.288 & 0.061 \\ \cline{7-10} 
\raisebox{2.5ex}[0pt]{18} & \raisebox{2.5ex}[0pt]{$(13,25)$} & \raisebox{2.5ex}[0pt]{8.226} & \raisebox{2.5ex}[0pt]{-4.118} & \raisebox{2.5ex}[0pt]{16.987} & \raisebox{2.5ex}[0pt]{0.462} & 1.984&0.667&0.293& 0.062\\ \hline \hline

  &  &  &   &   & &  2.408 & 1.131 & 0.425 & 0.085 \\ \cline{7-10} 
\raisebox{2.5ex}[0pt]{20} & \raisebox{2.5ex}[0pt]{$(16,24)$} & \raisebox{2.5ex}[0pt]{-0.021} & \raisebox{2.5ex}[0pt]{-3.879} & \raisebox{2.5ex}[0pt]{17.862} & \raisebox{2.5ex}[0pt]{0.437} & 2.419&1.157&0.43774& 0.088\\ \hline \hline

  &  &  &   &   & &  2.284 & 0.624 & 0.334 & 0.058 \\ \cline{7-10} 
\raisebox{2.5ex}[0pt]{22.5} & \raisebox{2.5ex}[0pt]{$(15.5,29.5)$} & \raisebox{2.5ex}[0pt]{-9.994} & \raisebox{2.5ex}[0pt]{-4.064} & \raisebox{2.5ex}[0pt]{18.971} & \raisebox{2.5ex}[0pt]{0.435} & 2.490 &0.615&0.338& 0.0848\\ \hline \hline

  &  &  &   &   & &  3.310 & 1.092 & 0.514 & 0.091 \\ \cline{7-10} 
\raisebox{2.5ex}[0pt]{25} & \raisebox{2.5ex}[0pt]{$(21,29)$} & \raisebox{2.5ex}[0pt]{-20.602} & \raisebox{2.5ex}[0pt]{-4.050} & \raisebox{2.5ex}[0pt]{19.982} & \raisebox{2.5ex}[0pt]{0.391} & 3.648&1.426&0.598& 0.105\\ \hline \hline

\end{tabular}
\end{center}
\label{ppSDE}
\end{table}

At this point we have in hand, estimates of the constant
and linear sensitivities of the
coefficients of the SDE.
 The ``global" drift and diffusion coefficient
 functions 
  can be approximated by the following interpolation procedure:
\[
f_j(x)=\frac{w^L f_j^{L}(x)+w^R f_j^{R}(x)}{Z}
\]

where $j=1,2$ corresponds
 to the drift and diffusion
 coefficients respectively,
 $w^L$ and $w^R$ are the
 weights associated
 with the left and right
 expansion point, $f_j^{L}(x)$
 is the affine approximation to
 the nonlinear function based on
  the closest expansion point
   whose value is less than or equal to $x$
   (similarly for $f_j^{R}(x)$, but use the
    nearest expansion point strictly greater than $x$).
     The weights were assigned by the \emph{ad hoc} rule:

\[
w^L(x)=\frac{\left(\sigma_C+\sigma_D(x)\left(x-E_i\right)\right)^{-1}(E_{i+1}-x)}{E_{i+1}-E_i}
\]
\[
w^R(x)=\frac{\left(\sigma_C+\sigma_D(x)\left(E_{i+1}-x\right)\right)^{-1}(x-E_i)}{E_{i+1}-E_i}
\]
\[
Z=w^L+w^R
\]

The results of this simple interpolation procedure
are shown as a dotted lines in figure \ref{f_CIRpieces}
(in the rule above, $\sigma_C$ and $\sigma_D$ are the
empirically measured parameter standard deviations of
 the constant and linear term of the diffusion term;
  the interpolation rule for the drift is analogous).  Figure \ref{f_CIRpieces} 
  plots the relative error of using this
  procedure using the known SDE coefficient functions
  as the ``truth".  The errors in the diffusion coefficient function do not indicate a systematic
  bias in contrast to the drift coefficient \footnote{I offer an intuitive explanation for this fact;
  if one \emph{naively assumes} that the true paths come from an Euler
  simulation with Euler step sizes corresponding to
   the observation frequency then one is in the
   ``Gaussian case".  In between expansion points, the
   Taylor expansion of the known diffusion function
   consistently over estimates the diffusion function at nearby points ( the diffusion coefficient is  concave).  The \emph{true} transition density of the SDE has
     a smaller variance (the data generating process) than the deterministic Taylor series prediction. When one uses the simple Euler density approximation  in QMLE, the magnitude of the mean reversion parameter appears
       to be larger than it really is because of the error in the Taylor series approximation of the diffusion  function (larger noise is expected making the deterministic trend appear stronger).
       The actual situation is much
       more complicated due to the fact
       that the distributions are not Gaussian, a
        complicated nonlinear function is used to
         obtain parameter estimates (QMLE), finite sampling effects, etc.;
          however figure \ref{f_CIRpieces}
          is consistent with this overly simplified
          intuitive explanation.
          \label{cireulexplanation}}.

\begin{figure}
\includegraphics[angle=0,scale=.50]{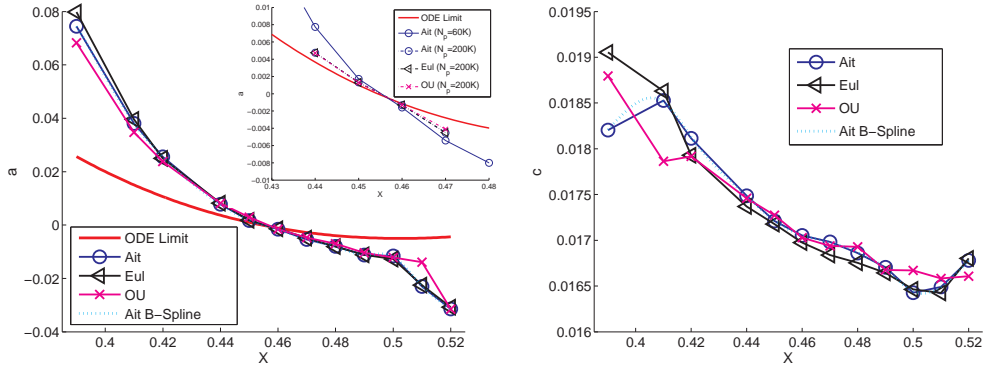}
\caption{Estimated Nonlinear Drift (Global) \small{
\emph{The LAQ screening method was used to obtain the  plots shown.  The results of three different transition density estimates are shown (the OU model sets $d$ in equation \ref{eq_myaffine} equal to zero). The left panel contains the drift coefficient corresponding to a SSA simulation containing $60^2$ particles. The thick solid line corresponds to infinite system size  drift function.  The dotted line corresponds to the B-spline of the \ait Expansion.  The inset   shows the drift coefficient corresponding to a SSA simulation containing $2\times 10^5$ particles.  Observe how the drift function convergence towards the expected limit equation with ``larger" systems, however for small particle systems the results are substantially different. The right panel displays the measured diffusion coefficient (the infinite sample limit has zero noise)      }}
}

\label{f_driftKMC}
\end{figure}

         In the final application, the parameters of the model are unknown.   The parameter were obtained with the LAQ screening technique.  The SSA process was run until $N$ sample paths reached  approximately ``invariant distributions" \footnote{  It is known that the free energy surface of this model contains two stable free energy wells and one saddle; no particles were able to overcome the large free energy barrier for the time series lengths used, hence the quotes on ``invariant distributions" }.  Then $N=200$ paths were used to obtain initial parameter estimates; for expansion points in the tails of the ``invariant distribution" an additional $N=600$ paths were simulated in order to get sharper estimates of the poorly sampled state points \footnote{Parameters were initially optimized over individual paths in order to estimate the parameter distribution variance.  When parameter were optimized on a pathwise basis, a significant fraction ($\approx 25\%$) of observation pairs resulted in \emph{assumed} spurious singularities in the \ait expansion.  To constrain the parameter space explored in the optimization,  I found the QMLE parameters with \emph{all} of the data (over time and paths).  This helped prevent the optimization routine from attempting to evaluate the log likelihood function at parameter values that cause spurious singularities because the parameter space explored was reduced because the trial  QMLE parameters needed to be good for all of the paths.  }.  Here  a B-spline \cite{deboor} was used ( MATLAB's cubic smoothing spline, ``$\mathbf{\mathrm{spaps}"}$ was used) to piece together the constants $(a,c)$ of the local models in order to smoothly interpolate between state points.   Each point was given the same weight when creating the spline, the LAQ technique for measuring parameter uncertainty could be used to give different weights to the measured parameters, but this procedure was not carried out here because the inherent jump nature of the data complicates matters slightly \footnote{In practice one could overcome this difficulty by obtaining the QMLE,  generate SDE sample paths with the model parameters obtained, then find the LAQ parameter variance of a genuine diffusion.  In applications where the uncertainty associated with using the local SDE model technique on imperfect data is desired, the problem  is much harder. One should consult the specialized literature \cite{semipar,rieder,vandervaart} for guidance; this author can not make any sound general recommendations.  }. The information contained in the linear coefficients $(b,d)$ was not used for interpolation purposes because the variance of the parameter of the linear terms was much larger than those of the constant terms in the case studied.  However, if one ignores the linear terms in the local model the constants estimated are significantly affected (see figure \ref{f_driftKMC}).  One observes  significant departure of the SSA model parameters from the limiting drift function (plotted as a dashed red line).  The number of particles in the SSA system was increased to $2\times 10^5$ and the parameters of the effective model in the drift were estimated and demonstrate that the limiting drift function is measured with the estimators used (see inset in figure \ref{f_driftKMC}). As the system size increases, the noise decreases, limiting the state points visited (hence the drift function estimated does not cover as large as a range as the first SSA case considered).  The estimated diffusion function in figure \ref{f_driftKMC} demonstrates dependence on the transition density used (\ait order one and Euler).  In this application the Ornstein-Uhlenbeck (OU) model was also used to demonstrate the importance of accounting for state dependent noise (note the systematic difference in the constant noise parameter estimated using the three different transition density estimators).

\begin{figure}
\includegraphics[angle=0,scale=.65]{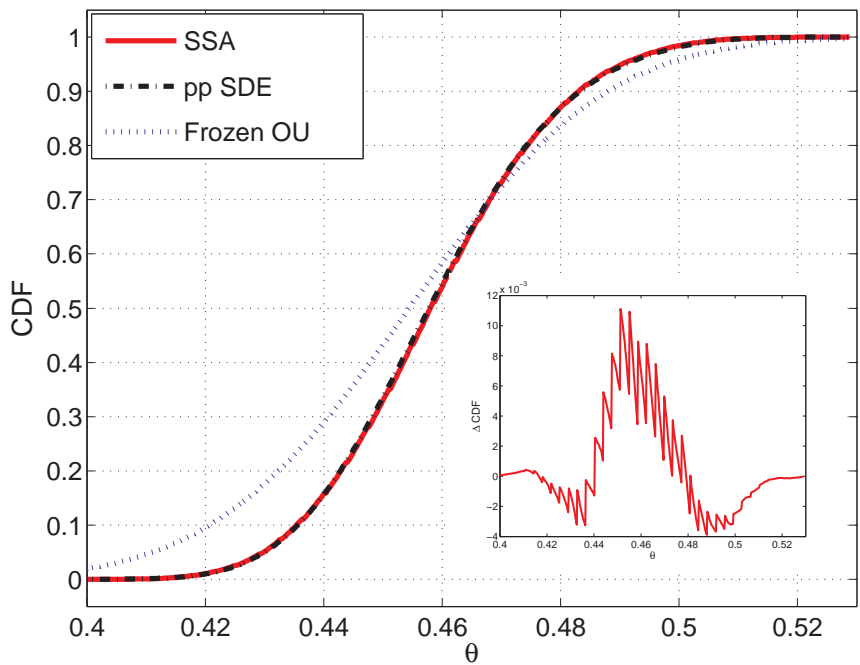}
\caption{CDF and KS test ($N_{molecules}=3600$) \small{
\emph{The CDF of the SSA ``invariant distribution" (empirically measured) plotted against that of the piecewise-polynomial (pp) local SDE model (simulated by long time Euler integration \cite{kp}).  In addition  the (actual) invariant distribution of the OU model obtained is plotted (the model parameters of the local OU model were measured at the \emph{single} state point where the drift is zero).  This was done to show what effect neglecting the nonlinear drift and state dependent diffusion has on the invariant CDF.  The inset plots the difference between the  ``invariant CDF" of the SSA data and that of the pp SDE model.  }} 
}

\label{f_cdfcomp}
\end{figure}
Figure \ref{f_cdfcomp}  plots the ``invariant" empirical CDF of the SSA data versus that predicted by a sample path simulation of our obtained nonlinear diffusion approximation (constructed from the B-spline interpolation of the estimated local SDE models).  The inset  plots the difference between the two empirical CDF's.  Our interest in this application was in getting a parametric description of the ``invariant" distribution, if the dynamics of the process are of more interest consult \cite{diebold} for a useful test which is made possible if one has an approximation of the transition density.           
          
\label{s_resgood}
\section{Conclusions and Outlook}
In this paper I have demonstrated that the expansions of A{\"\i}t-Sahalia  can potentially be a  useful tool in the parametric estimation associated with computational studies of  multiscale systems.  Parameter estimates can accurately be obtained and the curvature of the model is accurately represented  by the expansion in a variety of applications.  These facts can be exploited to estimate parameter distributions and construct useful inference procedures. The overly simple Euler expansion behaves poorly  in accuracy of the estimate and in the curvature of the transition density (as shown early on by Lo \cite{lo}), but it has the redeeming feature that it does not introduce any spurious singularities into the transition density (for the CIR model).  In large sample statistics applications, point singularities can \emph{usually} be remedied  \cite{lecam60,lecam86,lecam00} by using likelihood ratio expansions.  Unfortunately, many of these techniques require an extremely accurate estimate of the transition density. For imperfect transition densities with systematic errors, this becomes mildly problematic (from a computational standpoint) \footnote{ Methods proposed in \cite{aitextension}  are applicable to a wider class of models and  help the accuracy of the transition density expansion in the scalar case, but the vector case poses a more challenging problem}. 

I have also demonstrated a heuristic method for locally approximating a nontrivial SDE by a collection of simple local models.  The application was inspired by the need to accurately measure the parameters, quantify the uncertainty and determine the goodness-of-fit of parametric diffusion models around atomistic data.  The pp SDE technique presented is simple in nature, but it raises many deep questions.  Insight from the  semiparametric, robust and large sample statistics communities would greatly assist in further developing this type of numerical method.  The general method is appealing because it can be used to estimate nonlinear effective diffusion models where the effective SDE's coefficients are smooth, but of unknown functional form. The resulting parametric model structure (which is typically a complicated function due to the ``matching" used) can then be passed on to  diffusion path simulation methods in order to generate additional data which can be used in order to construct confidence bands or carry out established inference procedures.  

From a practical point of view it would be desirable to apply the techniques in this paper to A{\"\i}t-Sahalia's method  in the vector case because many interesting physical systems depend on a couple of ``reaction coordinates" \cite{hummer,yannisDIFFMAP,pittguy}.  Unfortunately the vector version of the expansion usually requires one to use an additional Taylor series approximation  (which increases the chance of spurious singularities and decreases the quality of the curvature estimate);  this researcher has been able to use the expansions in order to get useful parameter estimates, but has not had as much success in pushing them as far as the scalar expansions.  A numerical method which can be used in conjunction with the expansions of A{\"\i}t-Sahalia in order to get detailed information about the log likelihood ratio associated with smooth SDE's (diffusion and jump models) is currently being explored by the author.   

 \label{s_conclusions}

\section{Acknowledgements}
The author would like to thank Ioannis Kevrekidis and Adam Meadows for comments and advise and Yacine \ait for providing help with the transition density expansions.  Any errors are entirely due to the fault of the author.
\label{s_ack}

\bibliographystyle{siam}
\bibliography{LLglassyBIB}

\begin{thebibliography}{10}

\bibitem{kurtzbook}
{\sc S.~Ethier and T.~Kurtz}, {\em Markov processes : Characterization and
  convergence}, Wiley, 1986.

\bibitem{ait1}
{\sc Y.~A{\"\i}t-Sahalia}, {\em Transition densities for interest rate and
  other nonlinear diffusions}, Journal of Finance, 54 (1999), pp.~1361--1395.

\bibitem{ait2}
\leavevmode\vrule height 2pt depth -1.6pt width 23pt, {\em Closed-form
  likelihood expansions for multivariate diffusions}, Technical Report,
  (2001).

\bibitem{aitECO}
\leavevmode\vrule height 2pt depth -1.6pt width 23pt, {\em Maximum-likelihood
  estimation of discretely-sampled diffusions: A closed-form approximation
  approach}, Econometrica, 70 (2002), pp.~223--262.

\bibitem{aitextension}
{\sc G.~Bakshi and N.~Ju}, {\em A refinement to A{\"\i}t-sahalia's (2002)
  ``Maximum likelihood estimation of discretely sampled diffusions: A
  closed-form approximation approach"}, Journal of Business, 78 (2005), (to appear).

\bibitem{basawa}
{\sc I.~Basawa and {D.J.} Scott}, {\em Asymptotic Optimal Inference for
  Non-Ergodic Models}, Springer-Verlag, 1983.

\bibitem{bibby}
{\sc B.M. Bibby and M.~S{\o}rensen}, {\em Martingale estimation functions for
  discretely observed diffusion processes},Bernoulli, 1 (1995), pp.~17--39.

\bibitem{bickel}
{\sc P.~J. Bickel and K.J. Doksum}, {\em Mathematical Statistics: Basic Ideas
  and Selected Topics}, Prentice Hall, Upper Saddle River, NJ, 2001.

\bibitem{semipar}
{\sc P.~J. Bickel, C.~A.~J. Klaassen, Y.~Ritov, and J.~A. Wellner}, {\em
  Efficient and Adaptive Estimation for Semiparametric Models}, Johns Hopkins
  University Press, Baltimore, 1993.

\bibitem{deem}
{\sc {R.A.} Curtis and {M.W.} Deem}, {\em A statistical mechanics study of ring
  size, ring shape, and the relation to pores found in zeolites}, J. Phys.
  Chem. B, 107 (2003), pp.~8612--8620.

\bibitem{deboor}
{\sc C.~de~Boor}, {\em A practical guide to splines}, Springer, New York, 2001.

\bibitem{diebold}
{\sc {F.X.} Diebold, T.~Gunther, and A.~Tay}, {\em ''Evaluating density
  forecasts with applications to financial risk management"}, International
  Economic Review, 39 (1998), pp.~863--883.

\bibitem{paralleltemperingDEEM}
{\sc {D.J.} Earl and {M.W.} Deem}, {\em Optimal allocation of replicas to
  processors in parallel tempering simulations}, J. Phys. Chem. B, 1078 (2004),
  pp.~6844--6849.

\bibitem{pittguy}
{\sc B.~Ensing, A.~Laio, F.L. Gervasio, M.~Parrinello, and M.L. Klein}, {\em A
  minimum free energy reaction path for the E2 reaction between fluoro ethane
  and a fluoride ion}, JACS, 126 (2004), pp.~9492--9493.

\bibitem{yannisARXIV}
{\sc R.~Erban, {I.G.} Kevrekidis, D.~Adalsteinsson, and T.~Elston}, {\em Gene
  regulatory networks: A coarse-grained, equation-free approach to multiscale
  computation}, arXiv:physics/0508112,  (2005).

\bibitem{fan}
{\sc J.~Fan}, {\em Nonlinear time series: Nonparametric and parametric
  methods}, Springer, New York, 2003.

\bibitem{feigin}
{\sc P.~Feigin}, {\em Asymptotic theory of conditional inference for stochastic
  processes}, Stochastic Processes and their Applications, 22 (1986),
  pp.~89--102.

\bibitem{frenkel}
{\sc D.~Frenkel and B.~Smit}, {\em Understanding Molecular Simulation: From
  Algorithms to Applications}, Academic-Press, 2002.

\bibitem{gallant}
{\sc A.R. Gallant and G.~Tauchen}, {\em Which moments to match?}, Econometric
  Theory, 12 (1996), pp.~657--681.

\bibitem{lebowitz}
{\sc G.~Giacomin and J.L. Lebowitz}, {\em Phase segregation dynamics in
  particle system with long range interactions II: Interchange motion}, Siam J.
  Appl. Math., 58 (1998), pp.~1707--1729.

\bibitem{tau2}
{\sc {D.T.} Gillespie and {L.R.} Petzold}, {\em Improved leap-size selection
  for accelerated stochastic simulation}, J Chem. Phys. , 119
  (2003), pp.~8229--8234.

\bibitem{klaus}
{\sc J.~Gullingsrud, R.~Braun, and K.~Schulten}, {\em Reconstructing potentials
  of mean force through time series analysis of steered molecular dynamics
  simulations},J. Comp. Phys., 151 (1999), pp.~190--211.



\bibitem{halmos}
{\sc {P.R.} Halmos and {L.J.} Savage}, {\em Application of the {Radon-Nikodym}
  theorem to the theory of sufficient statistics}, The Annals of Mathematical
  Statistics, 20 (1949) , pp.~225--241.

\bibitem{hamilton}
{\sc {J.D.} Hamilton}, {\em Time Series Analysis}, Princeton University Press,
  1994.

\bibitem{schulten}
{\sc C.~Hardin, {M.P.} Eastwood, M.~Prentiss, Z.~Luthey-Schulten, and {P.G.}
  Wolynes}, {\em Folding funnels: The key to robust protein structure
  prediction}, J. Comp. Chem., 23  (2002), pp.~138--146.

\bibitem{hong}
{\sc {Y.} Hong and {H.} Li}, {\em Nonparametric specification testing for
  continuous-time models with applications to term structure of interest
  rates}, The Review of Financial Studies, 18  (2005), pp.~37--84.

\bibitem{hummer}
{\sc {G.} Hummer and {I.G.} Kevrekidis}, {\em Coarse molecular dynamics of a
  peptide fragment: Free energy, kinetics, and long-time dynamics
  computations}, J Chem. Phys., 118 (2003), pp.~10762--10773.

\bibitem{jeganathan}
{\sc P.~Jeganathan}, {\em Some aspects of asymptotic theory with applications
  to time series models}, Econometric Theory, 11 (1995), pp.~818--887.

\bibitem{yannisgear}
{\sc {I.G.} Kevrekidis, {C.W.} Gear, and {G.} Hummer}, {\em Equation-free: The
  computer-aided analysis of complex multiscale systems}, AIChE Jounral, 50
  (2004), pp.~1346--1355 .

\bibitem{kp}
{\sc P.~Kloeden and E.~Platen}, {\em Numerical Solution of Stochastic
  Differential Equations}, Springer-Verlag, 1992.

\bibitem{dima}
{\sc {D.I.} Kopelevich, {A.Z.} Panagiotopoulos, and {I.G.} Kevrekidis}, {\em
  Coarse-grained kinetic computations for rare events: Application to micelle
  formation}, J Chem. Phys. , 122 (2005), p.~044908.




\bibitem{kl}
{\sc S.~Kullback and {R.A.} Leibler}, {\em On information and sufficiency}, The
  Annals of Mathematical Statistics, 22 (1951), pp.~79--86.

\bibitem{stu}
{\sc R.~Kupferman and {A. M.} Stuart}, {\em Fitting SDE models to nonlinear
  Kac-Zwanzig heat bath models}, Physica D., 199 (2004), pp.~279--316.


\bibitem{pablo}
{\sc E.~La~Nave, F.~Sciortino, P~Tartaglia, {M.S.} Shell, and {P. G.}
  Debenedetti}, {\em Test of non-equilibrium thermodynamics in glassy systems:
  The soft-sphere case}, Phys Rev. E, 68 (2003), p.~032103.

\bibitem{lecam60}
{\sc L.~Le~Cam}, {\em Locally asymptotically normal families of distributions},
  University of California Publications in Statistics, 3 (1960), pp.~37--98.

\bibitem{lecam86}
\leavevmode\vrule height 2pt depth -1.6pt width 23pt, {\em Asymptotic Methods
  in Statistical Decision Theory}, Springer-Verlag, New York, 1986.

\bibitem{lecamrev}
\leavevmode\vrule height 2pt depth -1.6pt width 23pt, {\em Maximum likelihood:
  An introduction}, International Statistical Review, 58 (1990), pp.~153--172.

\bibitem{lecam00}
{\sc L.~Le~Cam and G.~L. Yang}, {\em Asymptotics in Statistics: Some Basic
  Concepts}, Springer-Verlag, 2000.

\bibitem{lehmann}
{\sc E.L. Lehmann}, {\em Testing statistical hypotheses}, New York, 1959.

\bibitem{lo}
{\sc {H.W.} Lo}, {\em Maximum likelihood estimation of generalized Ito
  processes with discretely sampled data}, Econometric Theory, 4 (1988),
  pp.~231--247.

\bibitem{makeev}
{\sc A.~Makeev, D.~Maroudas, and {I. G.} Kevrekidis}, {\em Coarse stability and
  bifurcation analysis using stochastic simulators: Kinetic Monte Carlo
  examples}, J. Chem. Phys.,116 (2002), pp.~10083--10091.

\bibitem{millar}
{\sc {P.W.} Millar}, {\em The minimax principle in asymptotic decision theory},
  Lecture Notes in Math.,  (1983), pp.~75--265.

\bibitem{nualart}
{\sc D.~Nualart}, {\em The Malliavin Calculus and Related Topics}, Springer,
  New York, 1995.

\bibitem{smle}
{\sc {A.R.} Pedersen}, {\em A new approach to maximum likelihood estimation for
  stochastic differential equations based on discrete observations},
  Scandinavian J. of Statistics, 22 (1995), pp.~55--71.

\bibitem{rieder}
{\sc H.~Rieder}, {\em Robust asymptotic statistics}, Springer-Verlag, New York,
  1994.
\bibitem{yannisDIFFMAP}
{\sc O.~Runborg, J.Krishnan, and {I. G.}~Kevrekidis}, {\em Bifurcation analysis
  of nonlinear reaction–diffusion problems using wavelet-based reduction
  techniques}, Comp. Chem. Eng., 28 (2004), pp.~557--574.

\bibitem{yanniscatalyst}
{\sc E.~Schutz, N.~Hartmann, {Y.} Kevrekidis, and R.~Imbihl}, {\em
  Microchemical engineering of catalytic reactions}, Catalysis Letters,
  54 (1998), pp.~181--186.

\bibitem{vandervaart}
{\sc A.~van~der Vaart}, {\em Asymptotic Statistics}, Cambridge University
  Press, 1998.

\bibitem{wahba}
{\sc G.~Wahba}, {\em Spline models for observational data}, SIAM, Philadelphia,
  1990.

\bibitem{white}
{\sc H.~White}, {\em Maximum likelihood estimation of misspecified models},
  Econometrica, 50 (1982), pp.~1--25.

\end{thebibliography}

\end{document}